%% file: SpinChainRMP.tex
\pdfoutput=1 
\documentclass[rmp,aps,reprint,amsmath,amssymb,longbibliography,twocolumn]{revtex4}
\usepackage{bm}

\usepackage{graphicx}
\usepackage{color}
\usepackage[english]{babel}

\newcommand{\nico}[1]{{\color{black} #1}}

\begin{document}

\title{\textit{Colloquium:} Atomic spin chains on surfaces}

\author{Deung-Jang Choi}
\affiliation{Centro de F{\'{\i}}sica de Materiales
CFM/MPC (CSIC-UPV/EHU), Paseo Manuel de Lardizabal 5, 20018 Donostia-San Sebasti\'an, Spain}
\affiliation{Donostia International Physics Center (DIPC), Paseo Manuel de Lardizabal 4, 20018 Donostia-San Sebasti\'an,Spain}
\affiliation{Ikerbasque, Basque Foundation for Science, 48013 Bilbao, Spain}
\author{Nicolas Lorente}
\affiliation{Centro de F{\'{\i}}sica de Materiales
CFM/MPC (CSIC-UPV/EHU), Paseo Manuel de Lardizabal 5, 20018 Donostia-San Sebasti\'an, Spain}
\affiliation{Donostia International Physics Center (DIPC), Paseo Manuel de Lardizabal 4, 20018 Donostia-San Sebasti\'an,Spain}
\author{Jens Wiebe}
\affiliation{
Department of Physics, University of Hamburg, D-20355 Hamburg, Germany}
\author{Kirsten von Bergmann}
\affiliation{
Department of Physics, University of Hamburg, D-20355 Hamburg, Germany}
\author{Alexander F. Otte}
\affiliation{
Department of Quantum Nanoscience, Kavli Institute of Nanoscience, Delft University of Technology, Lorentzweg 1, 2628 CJ Delft, The Netherlands}
\author{Andreas J. Heinrich}
\affiliation{Center for Quantum Nanoscience, Institute for Basic Science (IBS), Seoul 03760, Republic of Korea}
\affiliation{Department of Physics, Ewha Womans University, Seoul 03760, Republic of Korea}

\begin{abstract}
Magnetism at low dimensions is a thriving field of research with exciting
opportunities in technology.  In the present \textit{Colloquium}, we focus
on the properties of 1-D magnetic systems on solid surfaces. From the
emulation of 1-D quantum phases to the potential realization of Majorana edge states,
spin chains are unique systems to study. The advent of scanning tunnelling microscope (STM) based
techniques has permitted us to engineer spin chains in an atom-by-atom fashion via atom manipulation and to
access their spin states on the ultimate atomic scale. Here, we present the current state of research on 
spin correlations and dynamics of atomic spin chains as studied by the STM. After a brief review of the
main properties of spin chains on solid surfaces, we classify spin chains
according to the coupling of their magnetic moments with the holding
substrate.  This classification scheme takes
into account that the nature and lifetimes of the spin-chain excitation
intrinsically depend on the holding substrate.  We first show the interest
of using insulating layers on metals, which generally results in an increase in the spin state's 
lifetimes such that their quantized nature gets evident and they are individually accessible.
Next, we show that the use of semiconductor substrates promises additional control through the tunable electron density via doping. 
When the coupling to the substrate is increased for spin chains on metals, 
the substrate conduction electron mediated interactions can lead to emergent exotic phases of the coupled spin chain-substrate conduction electron system. 
A particularly interesting example is
furnished by superconductors.  Magnetic impurities induce states in
the superconducting gap.  Due to the extended nature of the spin chain,
the in-gap states develop into bands 
that can  lead to the emergence of 1-D topological superconductivity and, consequently to the appearance of Majorana edge states.
Finally,
we give an outlook on the use of spin chains in spintronics,
quantum communication, quantum computing, quantum simulations and quantum
sensors. 

\end{abstract}


\date{\today}

\maketitle
\tableofcontents


\input{./INTRODUCTION/introduction}

\input{./THEORY/theory}

\input{./EXPERIMENTAL/experimental}

\input{./SpinChainsDecoupled/passivated}

\input{./SpinChainsDecoupled/kondo}

\input{./SpinChainsDecoupled/semiconductors}

\input{./SpinChainsCoupled/metals_superconductors}

\input{./OUTLOOK/outlook}

\section*{Acknowledgments}

D.-J.C. and N.L. acknowledge financial support from MINECO
(MAT2015-66888-C3-2-R) and FEDER funds. 
J.W. and K.v.B. have received funding from 
the Deutsche Forschungsgemeinschaft (DFG, German Research foundation) - SFB668.
 A.F.O. acknowledges support from
the Netherlands Organisation for Scientific Research (NWO) and from the
European Research Council (ERC Starting Grant 676895 'SPINCAD'). 
A.J.H
acknowledges support from Institute for Basic Science under IBS-R027-D1.

\bibliography{./spin_chain}

\end{document}

%% file: INTRODUCTION/introduction.tex
\section{Introduction}
\label{intro}

A collection of local magnetic moments arranged in a linear fashion
that interact via some spin-spin coupling is generally known as
a spin chain. This seemingly simple object is one of the most complex
and rich physical systems that have been studied since the advent of
quantum mechanics without a decline in interest ever since. As early
as 1928, Werner Heisenberg explained ferromagnetism using Pauli's
exclusion principle and the interaction between spins that bears his
name. Subsequently, antiferromagnetism was addressed in spin chains by the
seminal works of Bethe~\cite{Bethe} and Hulth\'{e}n~\cite{Hulthen_1938}. Also
in recent times the interest in spin chains continues.  The 2016
Nobel Prize explicitly mentioned spin chains through the work of
Haldane~\cite{Nobel,Haldane_1983}, that revolutionized the understanding
of condensed-matter physics by finding new phases of matter associated
to a certain set of the two interactions defining the spin-chain
parameters~\cite{Nobel,Haldane_1983,aff1989}. Additionally, the study of
spin chains has been instrumental in ushering the far-reaching concepts
of topology in condensed matter.

Spin chains are also paradigmatic integrable systems. Bethe developed
the Bethe Ansatz to solve antiferromagnetically coupled spin
chains~\cite{Bethe,Hulthen_1938}, which has found use in many other
integrable models~\cite{Sklyanin,Takhtajan,Faddeev}.

The simplification of spin chains as compared to three dimensional
systems, brings in new phenomena proper to lower dimensions. One of them
is the absence of long-range order as descibed by the Mermin-Wagner
theorem~\cite{Mermin-Wagner}.  A related consequence is that  phase
transitions in one-dimensional (1-D) systems only take place at zero Kelvin.
Furthermore, correlations are enhanced at 1-D. As a consequence, many-body
physics is ubiquitous in 1-D system.

\nico{
While the initial interest in spin chains was primarily from a theoretical
viewpoint, various ways exist to create physical realizations of spin
chains in either solids, trapped atoms or molecules. Particularly the development
of the scanning tunneling microscope (STM) has furthered permitted us
to create spin chains on solid surfaces with atomic precision. 
}

\nico{
The first experimental realizations of spin chains date from the early 1960s.  It was found
that some transition-metal salts had their magnetic centers arranged
in a chain-like fashion and showed exchange interactions between these
centers~\cite{Watanabe_1958,Haseda_1961,Flippen_1963,Wagner_1964}.
An interesting family of 1-D spin systems are based on Cu
ions~\cite{Sahling_2015}.  
Recent activity is moving instead into
the creation of extraordinary spin chains using molecular systems~\cite{Caneschi2001May,Review_MolChains}.
}

\nico{
A great deal of progress in the experimental investigation of the physics of spin chains has been achieved in developing quantum simulators
based on atomic traps. Spin interactions can be simulated
by the close-ranged interactions between atoms held in an optical trap~\cite{simon_2011}.
When strongly interacting multicomponent gases are arranged
in one dimension, effective Heisenberg spin chains can be
modelled~\cite{Deuretzbacher}.  Short-ranged strong interactions
between alkali atoms have also been used to simulate the Heisenberg XXZ
models~\cite{Volosniev,Yang}.
 More recently, simulations of antiferromagnetic Heisenberg spin chains
have been performed using four fermionic atoms~\cite{Murmann_2015}.
Also Yang et al.~\cite{Yang_2016} have shown that these floating atoms
can lead to other interesting examples of Heisenberg chains. 
}

\nico{
The present \textit{Colloquium}  is devoted to the study of spin
chains on solid surfaces as studied with the STM. The STM allows
us  to interrogate matter on the atomic scale with
unprecedented precision. Besides studying spin chains
built by self-assembling techniques, the STM can actively 
 displace,
transfer and position atoms on a solid surface~\cite{eigler_1990}.
In this way, spin chains can be built with atomic precision of
both the chain itself as well as its environment. 
Furthermore, recent progress has permitted to greatly enhance the
applications of STM by conferring it with the ability of measuring
single-atom magnetic excitations~\cite{heinrich_2004}, single-atom magnetisation curves~\cite{Meier_2008}, 
single-atom fast
time-resolved spin dynamics~\cite{Loth_2010}, and single-atom electron
paramagnetic resonances~\cite{Baumann_2015}.  With these new capabilities,
the spin chains can be assembled and characterized atom-by-atom with
unique combination of control and accuracy.  As a consequence a new world of
data is booming in the field of spin chains.
}

\nico{
The recent years have seen a great deal of activity in the
field of spin chains on solids. We review this activity
classifying
the STM-based research by the substrate system. This
allows us to review processes as interesting as Kondo physics in
heterogeneous spin chains~\cite{Choi_2017c} or as ground-breaking as
the observation of indications for Majorana edge states in condensed matter~\cite{Ali2014, Kim_2018}.
}

%% file: THEORY/theory.tex
\section{Concepts of spin chain physics}
\label{atomic}

The extraordinary interest in spin chains stems from their complex quantum
nature. In this section, we review the properties of spin chains by first
deriving simplified Hamiltonians that only consider 
 effective interactions among magnetic moments. Next, we study the excitation spectra
of these effective Hamiltonians, first by considering the Heisenberg
model and then the effect of magnetic anisotropy. Finally, we analyse
the complexity of these solutions by revealing the role of entanglement,
comparing it with many-body correlations and explaining the effect of
decoherence of spin chains on solid surfaces.

\subsection{Spin Hamiltonians}
\label{Spin_Ham}

The Hamiltonian of a non-relativistic atomic system representing, for instance, a
condensed-matter realization of a chain of spins, does not contain any
spin operator because the spin is contained in the electronic states. As a
consequence, the total spin operator ($\hat{S}^2$) and one of the components of the spin
(say $\hat{S_z}$) will commute with the Hamiltonian.  When relativistic effects
are included, the spin operator fully appears in the spin-orbit coupling
terms, and both $S$ and $S_z$ can cease to be good quantum numbers.

\textit{Heisenberg Hamiltonian. ---} Spin operators naturally appear in a Hamiltonian if we simplify matters
to only include the low-energy excitations of the full system. Generally,
magnetic excitations are of low energy and a spin Hamiltonian will
explicitly consider them. 

\nico{
Open-shell atoms have two sources of magnetic moment, $\hat{\vec{L}}$ and $\hat{\vec{S}}$
that add to give the magnetic moment 
$\hat{\vec{M}}=-\mu_B (\hat{\vec{L}}+2 \hat{\vec{S}})$.
}

Here, we will restrict ourselves to spins in a vague way, but they can be any
of the above operators that contribute to the magnetic moment of the
system. The aim of the spin Hamiltonian is to simplify the description of the magnetic
structure of the system by using effective interactions among spins. The Heisenberg
Hamiltonian is a clear case of a spin Hamiltonian. \nico{It is a simple model for
the interaction between two magnetic moments.} 
The actual interaction behind electrons giving rise to the effective
interaction can be quite intricate.  Take for example the superexchange
interaction between two localized orbitals, $1$ and $2$ (see~\cite{Yosida}
for more details). The original Hamiltonian only includes a nearest
neighbor hopping term, $t$, that leads to chemical hybridization,
and an on-site Coulomb repulsion term, $U$, that adds a penalty to
double occupations of some local orbitals. The low-energy excitations
can be represented by the solutions of a Heisenberg Hamiltonian
with an antiferromagnetic interaction given by 
\begin{equation}
\hat{H}_\mathrm{Heisenberg}= J \hat{\vec{S}}_1 \cdot \hat{\vec{S}}_2.  
\label{Heisenberg}
\end{equation} 
The coupling term is given by~\cite{Yosida}
\begin{equation} 
J=\frac{2 t^2}{U}.  
\label{J} 
\end{equation}

\nico{
In order to take into account the varying nature
of different 
localized magnetic moments,
we can generalize the Heisenberg Hamiltonian to:
\begin{equation}
\hat{H} = \frac{1}{2}  \sum_{ij} \hat{\vec{S}}_i \cdot \mbox{J}_{ij}\cdot \hat{\vec{S}}_j
\label{GH}
\end{equation}
with a full magnetic exchange tensor, $\mbox{J}_{ij}$, that takes into acount all 
couplings between different pairs of effective spins $\hat{\vec{S}}_i$, $\hat{\vec{S}}_j$ of localized magnetic moments $i$ and $j$.
}

\nico{
This operator can be separated into three contributions~\cite{arXiv1811}:
}
\begin{eqnarray}
\hat{H}&=& \underbrace{\frac{1}{2} \sum_{i\neq j} J_{ij}\hat{\vec{S}}_i
\cdot \hat{\vec{S}}_j}_{\text{isotropic exchange}} +
\underbrace{\frac{1}{2} \sum_{i \neq j} \vec{D}_{ij}
\cdot \left( \hat{\vec{S}}_i \times \hat{\vec{S}}_j
\right)}_{\text{Dzyaloshinskii-Moriya interaction}} \nonumber \\
  &+&
\underbrace{\frac{1}{2}  \sum_{i\neq j} \hat{\vec{S}}_i \cdot
\mbox{J}_{ij}^\text{aniso} \cdot\hat{\vec{S}}_j}_{\text{anisotropic
exchange}}
\label{Jens}
\end{eqnarray}
The above tensor of exchange interactions, $\mbox{J}_{ij}$, was split
into its constituents: the isotropic exchange interaction $J_{ij}$,
the Dzyaloshinskii-Moriya interaction (DMI) $\vec{D}_{ij}$, and the
symmetric anisotropic exchange interaction $\mbox{J}_{ij}^\text{aniso}=
(\mbox{J}_{ij}+(\mbox{J}_{ij})^\text{T})/2-J_{ij}$.

\nico{
The DMI can arise when the inversion symmetry
of a system with sizeable spin-orbit coupling is broken,
becoming one source of non-collinear arrangements of
spins~\cite{Dzyaloshinskii,Moriya,Fert1,Fert2}. The DM vector,
$\vec{D}_{ij}$, gives the strength and orientation of the interaction
and is subject to symmetry selection rules; this interaction minimizes
the energy for an orthogonal orientation of adjacent spins and
dictates the rotational sense of the spin vectors. In competition
with the isotropic and anisotropic Heisenberg exchange, 
it can lead to ground states that are spin spirals
exhibiting a unique rotational sense~\cite{MenzelPRL2012,Schwefinghaus}.  
The DMI is also an important ingredient
for the formation of magnetic skyrmions in two dimensions~\cite{Heinze}.
}

\nico{
All the different parts of the exchange interactions in equation~\ref{Jens} can in principle not only result from the super exchange 
discuss above, but also from direct exchange, for close distance between the two localized spin-carrying orbitals, or from conduction electron mediated
indirect exchange interaction for a further separation of the localized orbitals.
The isotropic part of the latter
type of interaction is known as the Ruderman-Kittel-Kasuya-Yosida
(RKKY) interaction~\cite{Ruderman,Kasuya,YosidaK}.  It typically
has a damped oscillatory dependence, \textit{i.e.} it changes between
ferromagnetic and antiferromagnetic coupling as a function of the
distance between two atomic spins and their orientation 
with respect to the substrate lattice. The latter behavior results from
the shape of the Fermi surface of the conductance electrons that
can be rather complex and anisotropic~\cite{Zhou2010}.  Because of the
inversion symmetry breaking due to the presence of a surface, 
the conduction electron mediated exchange interaction  also has a
Dzyaloshinskii-Moriya contribution that can be as large as the
isotropic contribution if the substrate electrons are subject to considerable
spin-orbit interaction~\cite{Smith,Fert1,Khajetoorians2016}.  As a result,
chains of indirect conduction electron exchange coupled atoms on \nico{high atomic number} metallic substrates can also have
spin-spiral ground states~\cite{Steinbrecher2017}.
}

\nico{
Finally, there can be higher-order terms of the exchange interaction. 
The next higher order involves hopping between four-spins, located on two, three, or four sites~\cite{Kurz,Bluegel,Hoffmann}.
}

\textit{Magnetic anisotropy. ---} \nico{So far, only the interactions between localized magnetic moments have
been considered in Eq.~(\ref{Jens}). However, the orbitals of the local
moment also interact with the surrounding ligands via Coulomb interactions.
Together with spin-orbit coupling, this leads to a dependence of the
system's energy on the orientation of the magnetic moment, the so-called on-site
magnetic anisotropy. In order to take this into account, 
an additional contribution is added to the Hamiltonian where
the orbital degrees of freedom of the electronic wavefunction are implicit
and only the spin degrees of freedom are actively considered.
For the sake of understanding we first consider a low-symmetry binding site which leads to sufficient splitting
of the orbital degrees of freedom.
}

\nico{
In the absence of spin-orbit interaction, if the value of the orbital angular momentum contribution to the magnetic moment
is negligible,
a particularly simple case takes place.
This is often the case when the symmetry is strongly broken by the 
substrate holding the magnetic atoms.
Due to this  quenching of the orbital magnetic moment,
the low-energy excitations of the effective spin Hamiltonian are free of active
orbital transitions
when the spin-orbit interaction is connected.
However,
despite the quenching of the orbital moment,
there will be a final non-zero value of the orbital moment due to the efficient mixing
of spin and orbital degrees of freedom by the spin-orbit interaction.
}

The lowest-order terms in the additional magnetic anisotropy contribution to the effective spin Hamiltonian correspond to uniaxial symmetry
of the ligand field
and, allowing for some non-trivial symmetry transversal to the main axis (such as $C_i$, $C_{s}$, $C_{2\nu}$, etc), have the form
\begin{equation}
\hat{H}= D\hat{S}_z^2+E (\hat{S}_x^2-\hat{S}_y^2).
\label{MAE}
\end{equation}
This Hamiltonian is found very often, as for example in the case of magnetic
impurities on Cu$_2$N surfaces as will be described in section~\ref{passivated}.
Other ligand or crystal symmetries lead to the survival or cancelling
of higher powers of the spin operators. The Stevens operators are a systematic way 
to include contributions to the spin Hamiltonian taking into account
the symmetry of the atomic environment~\cite{Stevens_1952,Rudowicz_2004}.
Stevens generalized the spin Hamiltonian to read:
\begin{equation}
\hat{H} = \sum_{k=2,4,6} \sum_{q=-k}^k B_k^q \hat{O}_k^q (\vec{S}).
\label{Stevens}
\end{equation}
Each of the $\hat{O}_k^q (\vec{S})$ operators is Hermitian and the
coefficients $B_k^q$ are real such that the spin Hamiltonian is
Hermitian. A rank $k$ of 6 is sufficient to describe the effects of
crystal-field symmetry on spins on surfaces.

The above axial anisotropy can be expressed using the Stevens
coefficients, $B_k^q$, for the above zero-field splitting parameters,
such that, \[D=3 B_2^0 \; \; \; \; \; \; E=B_2^2\] where $k=2$ implies that
they refer to axial symmetry and $q=0,2$ refer to the longitudinal
and transversal components respectively.  The corresponding Stevens
operators are: \[\hat{O}_2^0= 3 \hat{S}_z^2-S(S+1)\] where $S$ is
the spin eigenvalue and $\hat{S}_z$ is the $z$ component of the spin
operator. And, \[\hat{O}_2^2= \hat{S}_x^2-\hat{S}_y^2.\] The Stevens
operators are widely tabulated and can be found in many references,
for example~\cite{Rudowicz_2004} and references therein.

Many substrates possess a C$_{3\nu}$ symmetry. An example are
the substrates of Section~\ref{passivated}. In many instances, we will see
that higher order terms can be often neglected, such that using $D\hat{S}_z^2$ is
already good enough for those systems.

\begin{figure}
\includegraphics[width=1.0\columnwidth]{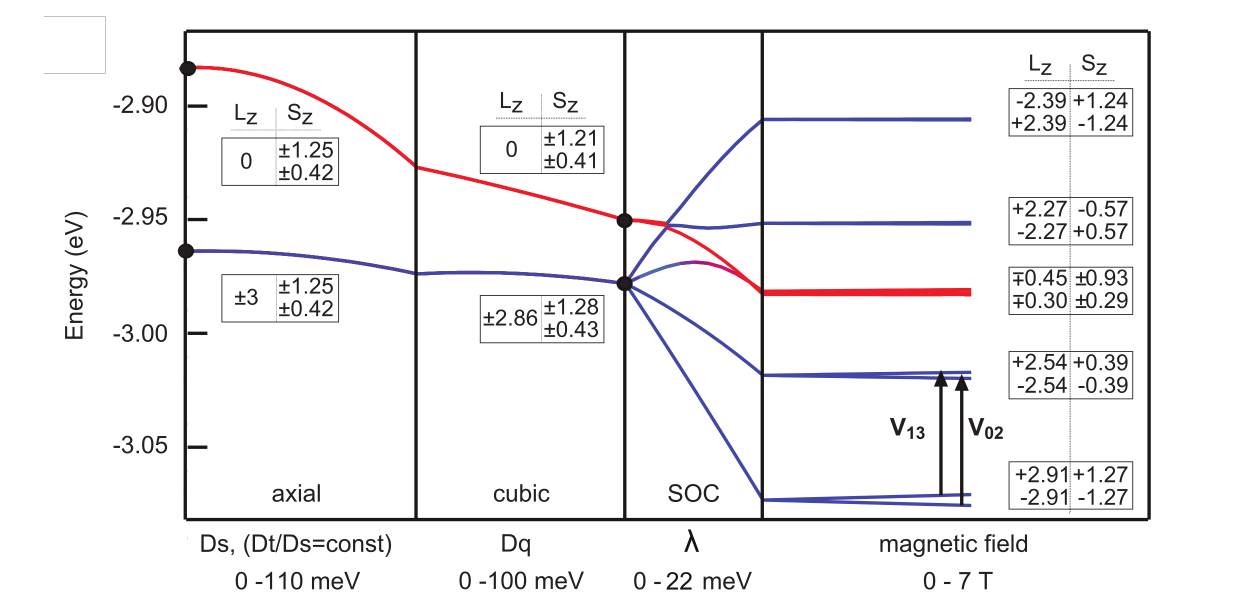}
\caption{\label{multiplete}
The effect of different perturbations on the electronic states of a Co$^{2+}$ ion on MgO is
incrementally shown in this figure. First the axial field due to the presence of
the surface plane is included, shifting the ten low-energy spherical levels (one eight-fold
and one two-fold degenerate), 
second the crystal field of the four neigboring Mg$^{2+}$ ions is considered. Next,
the spin-orbit coupling is adiabatically switched on. And finally a magnetic field is included.
The lowest energy transitions induced by tunneling electrons from an STM are depicted by
arrows. Reproduced from \cite{Rau_2014}.
}
\end{figure}

Writing such an effective spin Hamiltonian is not always possible. In
the absence of  quenching of the orbital degrees of freedom, the spatial dependence of the electronic
wavefunction has to be explicitly allowed in the Hamiltonian. This
case has been considered in a number of \nico{recent} works about
magnetic impurities on a MgO thin film on a Ag (100) substrate
\cite{Rau_2014,Baumann_2015,Baumann_2015b,Ferron_2015}.  These articles
study $3d$ transition metals on MgO layers. Generally, adsorption on one
of the surface's oxygen atoms is preferred, leading to an axial symmetry
given by the normal to the surface, plus a four-fold symmetry by the
four neighboring Mg$^{2+}$ ions. The axial crystal field is not strong
enough to sufficiently quench the orbital moment and a full multiplet
calculation must be undertaken. Figure~\ref{multiplete} shows the typical
procedure to obtain the low-energy terms of a Co$^{2+}$ ion on MgO. The
calculations proceed by first considering the axial field effect on
the electronic states of the studied $3d$ ion. Next, the four-fold
crystal field is added. Once that the electronic states reflect the
correct orbital structure under the environment's fields, the spin-orbit
interaction is added. Finally the Zeeman splitting due to an external
magnetic field is considered. This approach incrementally considers each
effect permitting us to gain insight as well as control on the actual
electronic configuration of the ion in its environment. By this procedure, a spin
Hamiltonian enhanced by orbital terms, similar to the above spin terms,
can be obtained that reproduces the low-energy states of the system.

\nico{
\textit{Classical magnetic moments. ---}
If the quantum properties of the spin chain system are not dominating,
\textit{e.g.} because of very large spin values, it is often sufficient
to consider the classical limit. Within this limit, the vector spin
operator $\hat{\vec{S}}$ is replaced by a classical magnetic moment,
via $\vec{m} = g \mu_B \vec{S}$, 
and the
parameters within the Stevens operator treatment are related to
classical magnetic anisotropy constants.
}

\subsection{Excitations in spin chains}
\label{single-particle}

The ground state of a ferromagnetic Heisenberg chain in the presence
of an arbitrarily small magnetic field corresponds to all spins
being aligned along the magnetic field. 
\nico{Flipping a spin does not result in an eigenstate of the Hamiltonian; instead, 
it forms a linear combination of eigenstates}~\cite{Yosida,Mattis,Auerbach}. Since the complete reversal of a single spin requires a lot of energy
due to the exchange interaction, the lowest-energy excitations of chains are
spin waves, where the spin flip is delocalized with a phase shift along
the entire chain. For an infinite chain of $S=1/2$ spins, a spin-wave excitation
has the following dispersion relation: 
\begin{equation} \epsilon
(\vec{q})= 2 J  \sin^2 (\frac{\vec{a} \cdot \vec{q}}{2}) 
\label{SW_ferro} 
\end{equation}
where $\vec{q}$ is the dispersion momentum vector along the infinite spin
chain and $\vec{a}$ \nico{the lattice vector of the spin chain}. Spin waves are also referred to
as magnons.

If the Heisenberg coupling between spins is instead antiferromagnetic, the spin
states are far from simple due to the multiconfigurational character
of the antiferromagnetic solutions.
 For chains of atoms with S=1/2, the flipping of one spin leads to
either a spin wave or a two-spinon excitation. \nico{Either of these excitation changes the total magnetization} by $\Delta S = \pm 1$. 
The spin wave is the lowest-energy
excitation of the antiferromagnetic chain and its dispersion relation
is given by \cite{desCloizeaux} 
\begin{equation} 
\epsilon (\vec{q})=
\frac{\pi}{2} J | \sin (\vec{a} \cdot \vec{q}) |, 
\label{desCloizeaux} 
\end{equation}
with the same notation as before.

Calculations on the probability and spectra of finite ferro- and
antiferromagnetic spin chains show that spin waves are efficiently
excited by tunneling electrons~\cite{Gauyacq_2011}.

\begin{figure}
\includegraphics[width=1.0\columnwidth]{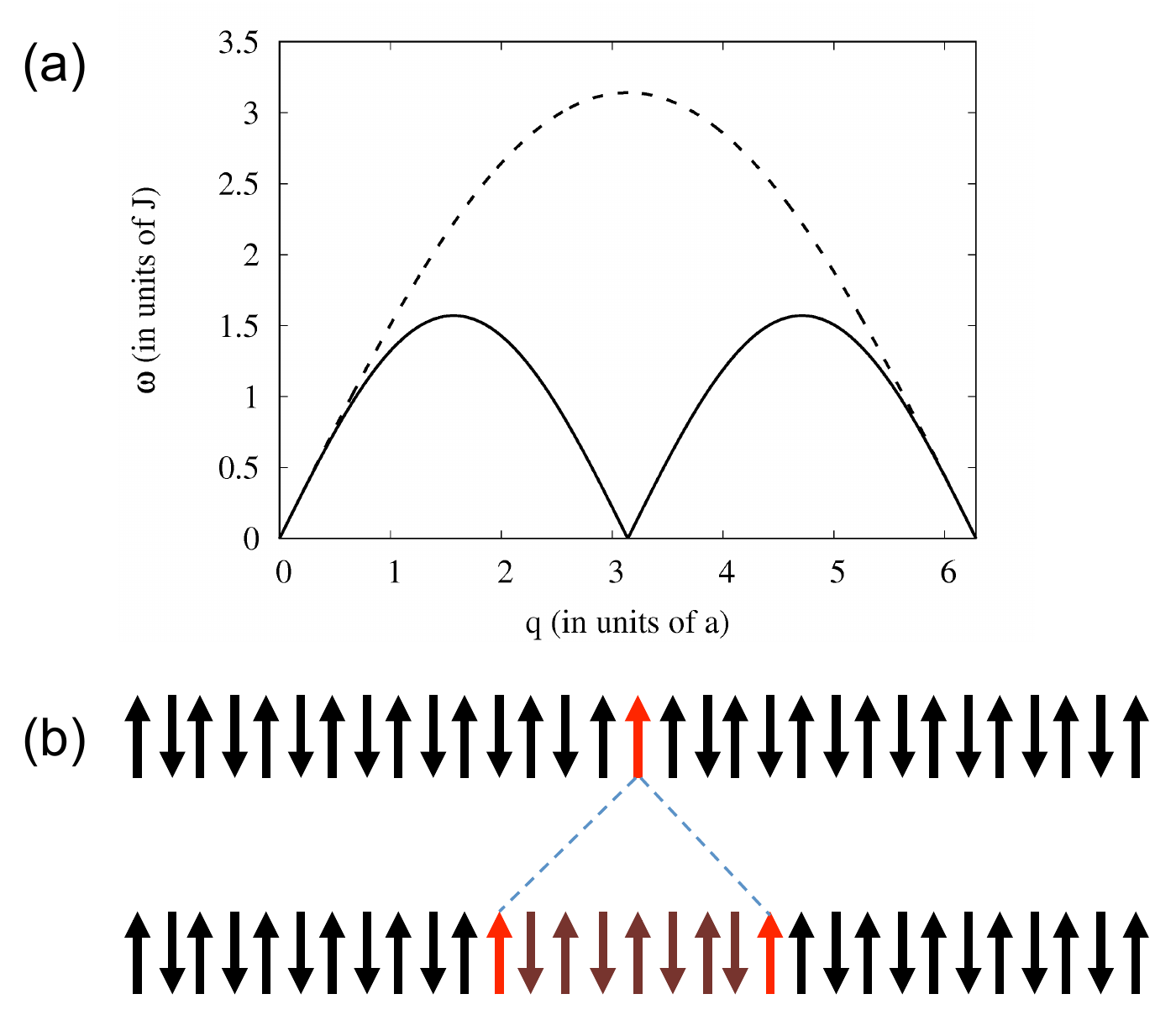}
\caption{\label{two-spinon}
(a) Two-spinon continuum corresponding
to single spin-flip excitations of an inifinite \nico{Heisenberg antiferromagnetic chain of atoms with S = 1/2 (spin 1/2 chain).}
The continuum is bounded by spinwave excitations as the low-energy branch, Eq.~(\ref{desCloizeaux}),
and the higher branch, Eq.~(\ref{upp}).
(b) Scheme of a spin 1/2 chain showing the propagation of a two-spinon
becoming two domain-wall excitations for an Ising antiferromagnetic chain.
}
\end{figure}

Right beyond the spin wave excitation two-spinon
excitations set in, until they reach the upper
boundary~\cite{Yamada_1969,Mueller_1981,Karbach_II,Karbach_1996,Kabach}.
\begin{equation}
\epsilon_U (\vec{q})= \pi J | \sin (\vec{a} \cdot \vec{q}/2) |.
\label{upp}
\end{equation}
Figure  \ref{two-spinon} (a)
shows the continuum of two-spinon excitations bounded by the
spin-wave excitation, Eq.~(\ref{desCloizeaux}), and the upper branch,
Eq.~(\ref{upp}). All these excitations  correspond to encountering one
spin flip in an antiferromagnetic spin chain. Half an excitation is a
spinon, which is a consequence of the tendency to the fragmentation of
spin (and charge) in 1-D systems \cite{Caux}.  This type of spectrum has
been recently revealed in 1-D spin chains formed by CuO \cite{Caux}. The
fragmentation of spin in the excited state is easily understood when the
time evolution of the two spinons is followed. Figure \ref{two-spinon} (b)
shows a simple scheme of the creation of a two-spinon excitation and its
time evolution into single spinons.  The two-spinon continuum is followed
by four-spinon excitations and so \nico{forth at even higher energies, but}~\cite{Mueller_1981,caux_2006}
most of the spectral weight is carried by the two-spinon
excitations~\cite{Mueller_1981,Karbach_1996}.

\subsection{Haldane phase}
\label{haldanephase}

For a while, it was believed that the above depicted excitation
spectra for spin 1/2 systems was general to all antiferromagnetic Heisenberg spin chains.
However, Haldane~\cite{Haldane_1983} predicted that the spectra for
 chains made out of integer spins ($S\ge 1$)
should be gapped, meaning that there are no zero-energy excitations
contrary to the spectra shown in Fig. \ref{two-spinon} (a).

Haldane conjectured
that the Heisenberg antiferromagnetic spin chain for integer spins
has a singlet ground state separated by an energy gap from the first
excited state~\cite{Haldane_1983}, see also \cite{aff1989,Tasaki_1991}.
This apparently minor difference has important implications.  The appearance of
the gap leads to spin-spin correlations that decay exponentially with
distance  while for the half-integer the spin-spin correlations decay
following a power law~\cite{Tasaki_1991,Renard_2002}. But moreover,
it leads to the possibility of non-trivial ground states for the
integer case with a corresponding topological quantum phase transition
between the different solutions~\cite{Wen,Pollmann_2010,Pollmann_2011,Pollmann}.  For a chain of $S = 1$ spins, the
Heisenberg antiferromagnetic spin chain with uniaxial anisotropy
($E=0$ in Eq.~(\ref{MAE})) presents a phase transition  for $D=J$
($J$ and $D$ defined in Eqs.~(\ref{Heisenberg}) and (\ref{MAE})).
Both states \nico{on either side of the phase transition} preserve all the symmetries of the Hamiltonian, hence the
phase transition does not take place by breaking symmetries, but it is
rather of topological nature~\cite{Chen2012}. The low $D$-phase is the Haldane phase
that is a strongly entangled state that cannot be smoothly connected
to a product state. However, the large $D$ phase can be connected to
 a simple product state. This last phase is the topologically trivial
one~\cite{Pollmann}.

\nico{The confirmations of the integer spin system being gapped are quite limited}, 
despite all the existing experimental work on 1-D spin
systems~\cite{Renard_1987,Renard_2002}.  Indeed, the requirements
to obtain the Haldane phase are somewhat difficult to find in a
physical system. The individual spins must be integers, the interaction
antiferromagnetic, arranged in 1-D periodical structures with uniform
interactions, but  weak interchain interactions and weak anisotropy. 
\cite{Renard_2002} give a complete list of Ni-based compounds with
quasi 1-D spin-one chains that present the Haldane phase.

An extension of Heisenberg spin chains is given by the AKLT model~\cite{AKLT}.
The AKLT model consists of a chain of sites that are connected by a bond.
This valence bond contains two spins 1/2. Then each site is effectively
a spin 1 system, but due to the valence bond that is singlet, the
sites are antiferromagnetically coupled. This model can be written into a spin
Hamiltonian by using projectors, arriving at the following expression:
\begin{equation}
\hat{H}=\sum_j (\hat{\vec{S}}_j \cdot \hat{\vec{S}}_{j+1} + \frac{1}{3} \hat{(\vec{S}}_j \cdot \hat{\vec{S}}_{j+1})^2)
\end{equation}
which is a spin 1 Heisenberg Hamiltonian plus an extra biquadratic term.
This model is exactly solvable, and its ground state
can be expressed as a matrix product state which still stirs a lot of theoretical attention.
Furthermore, the model has been used to study valence-bond order and
symmetry-protected topological order~\cite{Wen,Pollmann_2012}.

\subsection{Decoherence of spin chains}
\label{entanglement}

In the following, we will discuss decoherence effects
that arise by the interaction of the spin chain with the environment, which in this work is the holding substrate.

Let us first assume we have two $S=1/2$ spins  interacting via an exchange
interaction, $J \vec{S}_1 \cdot \vec{S}_2$.  We can diagonalize this
Hamiltonian and obtain three $S=1$ states and one $S=0$ state.  If we
measure one of the spins, we will find equal probabilities for spins
up and down. Thus,  we cannot obtain any information on the state
of an individual spin. However, we know the total spin of the two-spin
system. It is perfectly determined. The total states are \textit{canonical
maximally entangled} states~\cite{Hojckenczi}. Once we know the state
of the full system, and the state of one of the spins, we
will know the outcome of a possible measurement on the other spin.

This is true while the spins keep their respective relative phases. In
the events of collisions or perturbations that simply produce a change of
phase on one of the components, the wavefunction changes and the previous
entangled wavefunction does not represent the system anymore. Indeed,
for long enough times, the accumulation of dephasing events leads
to the collapse of the singlet wavefunction in either $|\uparrow
\downarrow\rangle$ or $|\downarrow \uparrow \rangle$ also known as
N\'eel states. When the spin chain is in contact with a substrate,
statistical interactions with the substrate (phonon or electron
collisions) lead to dephasing and hence decoherence.

In the case of spin chains on surfaces, the effect of decoherence
has been shown to lead to N\'eel-like solutions of antiferromagnetic
spin chains \cite{Gauyacq_2015,Delgado_2017}. 
It is instructive to compare
the cases of  Fe$_x$\cite{Loth_2012} and Mn$_x$\cite{Hir2006,Choi_2016} spin chains. 
 The main difference of these two systems is the magnetic anisotropy, Eq.~(\ref{MAE}).
The spin on Fe atoms on Cu$_2$N show a large anisotropy, while Mn displays a very small one.
As a consequence, the atomic spin of Fe has a preferential direction where it can easily align, and
create N\'eel-like states with aid from the environmental decoherence. Even for
similar decoherence rates, the absence of a preferential axis for Mn makes it more
difficult to collapse into a classical N\'eel state.

The time evolution of the density matrix
can be obtained  from the time evolution of the system state. 
The density matrix is an operator given by the projector on the state
of the full system, $|\Psi\rangle$, then the density
matrix is  $\hat{\rho}=|\Psi\rangle\langle\Psi|$. 
The time evolution leads to
\begin{equation}
\frac{d\hat{\rho}}{dt}=-\frac{i}{\hbar}[\hat{H},\hat{\rho}].
\label{timetot}
\end{equation}

Let us assume that we can express the total-system Hilbert space as the direct
product of the two subsystem Hilbert space: $\mathcal{H}=\mathcal{H}_A\otimes \mathcal{H}_B$,
where, for example, $A$ can stand for the spin chain and $B$ for the holding substrate. Once
we have determined $H_B$, we can use an eigenstate
basis,
\begin{equation}
\hat{H}_B |\phi_B\rangle=\epsilon_B|\phi_B\rangle,
\label{system_B}
\end{equation}
to project out the $B$ subsystem, leading to the reduced density matrix:
\begin{equation}
\hat{\rho}_A =\sum_B\langle \phi_B|\Psi\rangle\langle\Psi|\phi_B\rangle.
\label{reduced}
\end{equation}
When the reduced density matrix is used,
new terms explicitly appear in the time-evolution
equation, this can be written in terms of the dissipative part of the Liouvillian
$\mathcal {L}$\cite{Cohen}.
The actual way of doing this is very involved and many works
treat this problem~\cite{Cohen,Delgado_2017}. 

\begin{equation}
\frac{d \hat{\rho}_A}{d t} = - \frac{i}{\hbar} [ \hat{H}_A, \hat{\rho}_A]
+ \mathcal {L} (\hat{\rho}_A)
\label{timered}
\end{equation}

The effect
of the environment amounts to a source of random interactions between
the many degrees of freedom of the environment (subsystem B) and  the
degrees of freedom of subsystem A. 
The 
Liouvillian
can then be approximated by a linear term on the
differential equation for the coherences with a decay constant$\frac{1}{T_2^*}$.
Here, $\frac{1}{T_2^*}$ is the pure decoherence or pure dephasing rate.
Let us assume that we only have two states (1 and 2), then the dissipative part of the 
Liouvillian,
$\mathcal {L}$, is simply:
\begin{equation}
\mathcal {L} (\hat{\rho}_A)=-\frac{1}{T_2^*}\{\, \hat{ \rho}_{12}
+\hat{ \rho}_{21} \, \}.
\label{Liouvillian}
\end{equation}
Here, we have assumed no direct transition between states such
that $\frac{1}{T_1}=0$.
For more states, matrices have to be defined for the dephasing rates and the
equations become considerably more difficult without changing the physics.
A complete account of the quantum dynamics of a magnetic subsystem 
can be found in \cite{Delgado_2017}.

The above equations find direct application in the problem of the quantum
dynamics of a spin chain. The experiment by Loth \textit{et al.} consisted
in assembling an antiferromagnetic Fe chain on Cu$_2$N~\cite{Loth_2012}. The
spin-polarized STM image allowed them to measure the dwelling times in each
of the two
N\'eel state of the spin chain. They found that at very low
temperatures the switching rate between the two states was a constant with temperature. 

\begin{figure}
\includegraphics[width=0.55\textwidth]{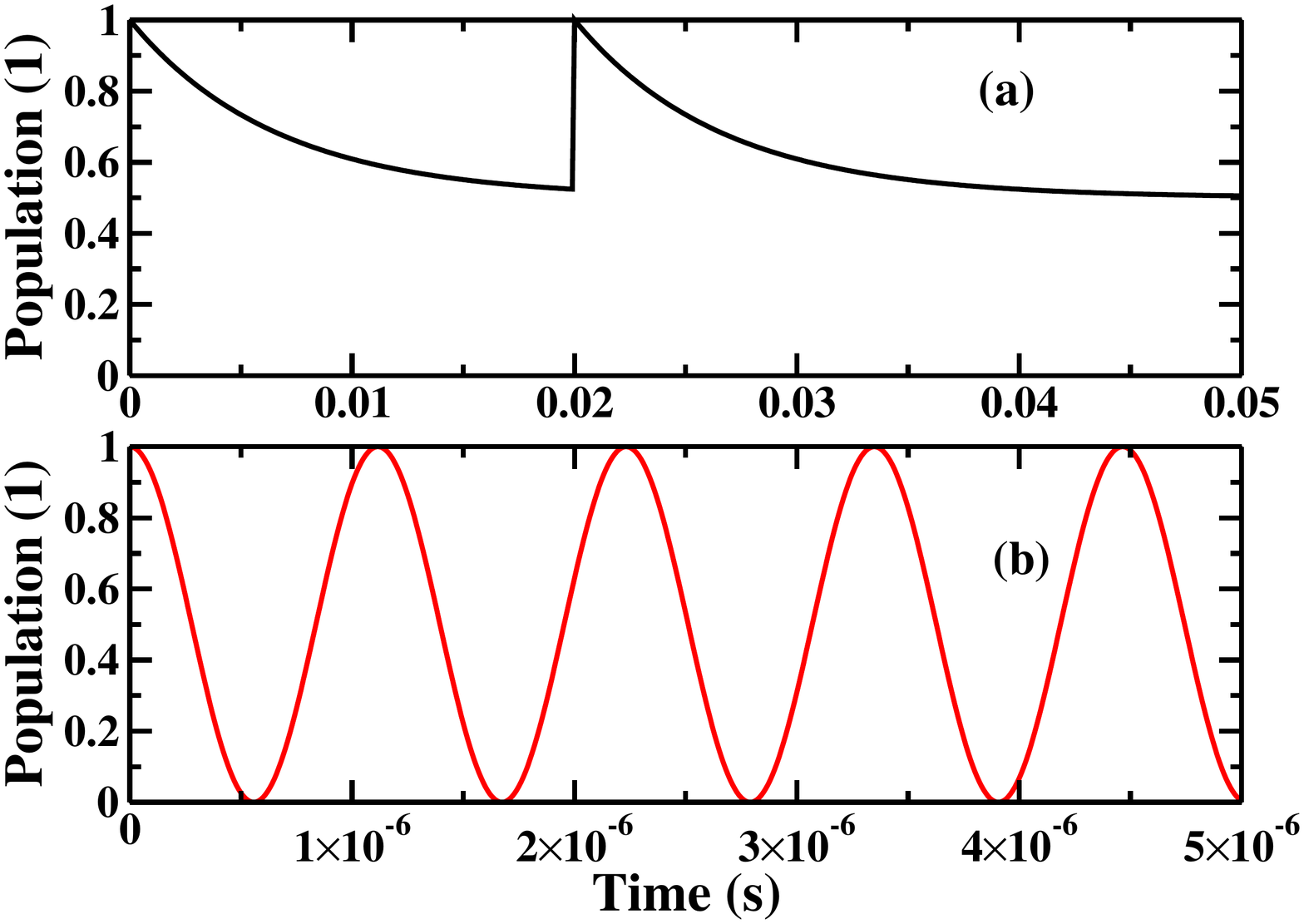}
\caption{
A {Fe$_6$} chain is initially in state 1 of the
two classical N\'eel states of this antiferromagnetic spin chain. 
In the presence of decoherence, the population of state 1 ($\rho_{11}$)
is exponentially reduced to 0.5, populating
both states, $(1)$, while the \textit{coherences} become zero ($\rho_{12}=\rho_{21}\rightarrow 0$).
Here, a measurement at 
 $\sim 0.02$ s is assumed to find the system in state 1,
then the population is suddenly 100\% for state 1. Afterwards
the exponential decay leads to 50\% populations.
In the absence of decoherence, the population
is given by Rabi oscillations of very fast frequency for the {Fe$_6$} chain
on Cu$_2$N. \textcolor{black}{Figure reproduced from~\cite{Gauyacq_2015}
with permission from the Institute of Physics.}
 \label{decoherence}}
\end{figure}

Calculations based on the above formalism showed that the spin-chain
dynamics was a competition between quantum tunneling, which leads to
Rabi oscillations between the two N\'eel states, and the decaying effect
of decoherence~\cite{Gauyacq_2015}. Pure quantum tunneling dynamics
leads to fast oscillations of the state populations.
However, due to decoherence,
the population evolution becomes exponential.  Figure~\ref{decoherence}
shows the difference between the spin-chain dynamics under decoherence,
$(a)$, or quantum tunneling alone, $(b)$. A factor of $10^{4}$ can be
rapidly gleaned from the time axis when comparing the time dependence of the
two curves.

%
%

%% file: EXPERIMENTAL/experimental.tex
\section{Experimental Methods and Sample Systems}
\label{experiment}

There are several experimental techniques typically used for the
preparation and investigation of spin chains on solid state
substrates depending on the nature of the spin chain and the properties
to study. Traditional methods are measurements over ensembles of many
spin chains and are thus averaging techniques such as, e.g., angle-resolved
photoemission spectroscopy, magnetic susceptibility, calorimetry, electron
spin resonance, and neutron scattering. The advent of scanning probe
techniques has permitted us to access to each individual atom in a
single spin chain. This gives rise to new possibilities such as studying
local properties by carefully positioning the scanning tunneling tip within
the spin chain, or studying chains as a function of number of atoms,
their nature and other parameters.  This section is devoted to a brief
description of the methods that can be used to study individual chains
regarding their magnetic properties, their preparation and the nature
of the holding substrate.

\subsection{Experimental Methods}

Here, we review the
methods  based on scanning
probe methods, particularly the STM. 
There are different STM measuring modes. The scanning modes typically
give information on the topography of the studied objects. For
spin chains they reveal important data on the atomic geometry
and disposition with respect to the substrate. The typical imaging mode
is the constant current mode where the set of tip heights over
the sample are recorded while keeping the tunneling current
constant. This very early measuring mode was shown to
largely reproduce the spatial distribution of the constant
local density of states (LDOS) of the substrate, at its
Fermi energy, $E_F$, and at the tip's
position~\cite{Tersoff}.

It was quickly realized that a second operation mode of the STM was to
measure the differential conductance at a given tip location. Extending
the interpretation of~\cite{Tersoff} to finite bias, $V$, this
would yield precious information on the density of states at a given
position, again the LDOS at the tip's location.  Furthermore, advanced
transport theory shows that in the presence of one conductance channel
or under some simplifying assumptions about the tip-substrate electronic
coupling~\cite{Meir}, the differential conductance is proportional to the
many-body LDOS of the substrate, at $E_F+eV$. Measuring the differential
conductance is tantamount to measuring the many-body spectral properties
of the substrate, ranging from any type of excitation, to the Kondo
effect and to the general orbital structure of the system. In summary,
the differential conductance contains information about all degrees of
freedom of the substrate that can interact with an injected electron.

The different ways to measure the differential conductance
give rise to different experimental techniques that we briefly review now:

\textit{Scanning tunneling spectroscopy.---}
\nico{In general, measuring the differential conductance at a given
bias $V$ and tip position is 
the spectroscopic mode named scanning tunnel spectroscopy
(STS). As we have just seen,
it provides information about the spin-averaged electronic properties
of a sample, and using a magnetic tip also spin-resolution is achieved
(see below).  Using the scanning capabilities of the STM,
maps of differential conductance can be produced at different
bias. When an interesting energy $E$ is identified spatially
resolved d$I$/d$V$~maps at only the according bias voltage $V=E/e$
can be performed to reduce the measurement time.
}

\textit{Inelastic electron tunneling spectroscopy.---}
The main experimental difference with the previous spectroscopic mode,
the STS, is the bias resolution that permits us
to obtain a direct measurement of inelastic excitations. 
In order to 
increase the signal-to-noise ratio compared to numerically
derived $dI/dV$ spectra 
lock-in techniques are applied.
The modulation should be high enough
to significantly reduce the $1/f$
noise, but low enough to be still in the bandwidth
of the amplifier. The modulation bias also
reduces the noise, at the expense of broadening
the spectral features.

These measuring mode
is usually known as inelastic electron tunneling spectroscopy (IETS).  Vibrational modes
ranging from a few to hundreds of meV, have been detected
with IETS~\cite{Stipe_1998,Ho_2002,Komeda_2005,Karina,morgenstern_2013}.
This was a very exciting development because IETS permitted
a chemical identification of adsorbed species that is generally
absent in the large-energy scale of STS.

\begin{figure}
\includegraphics[width=1.0\columnwidth]{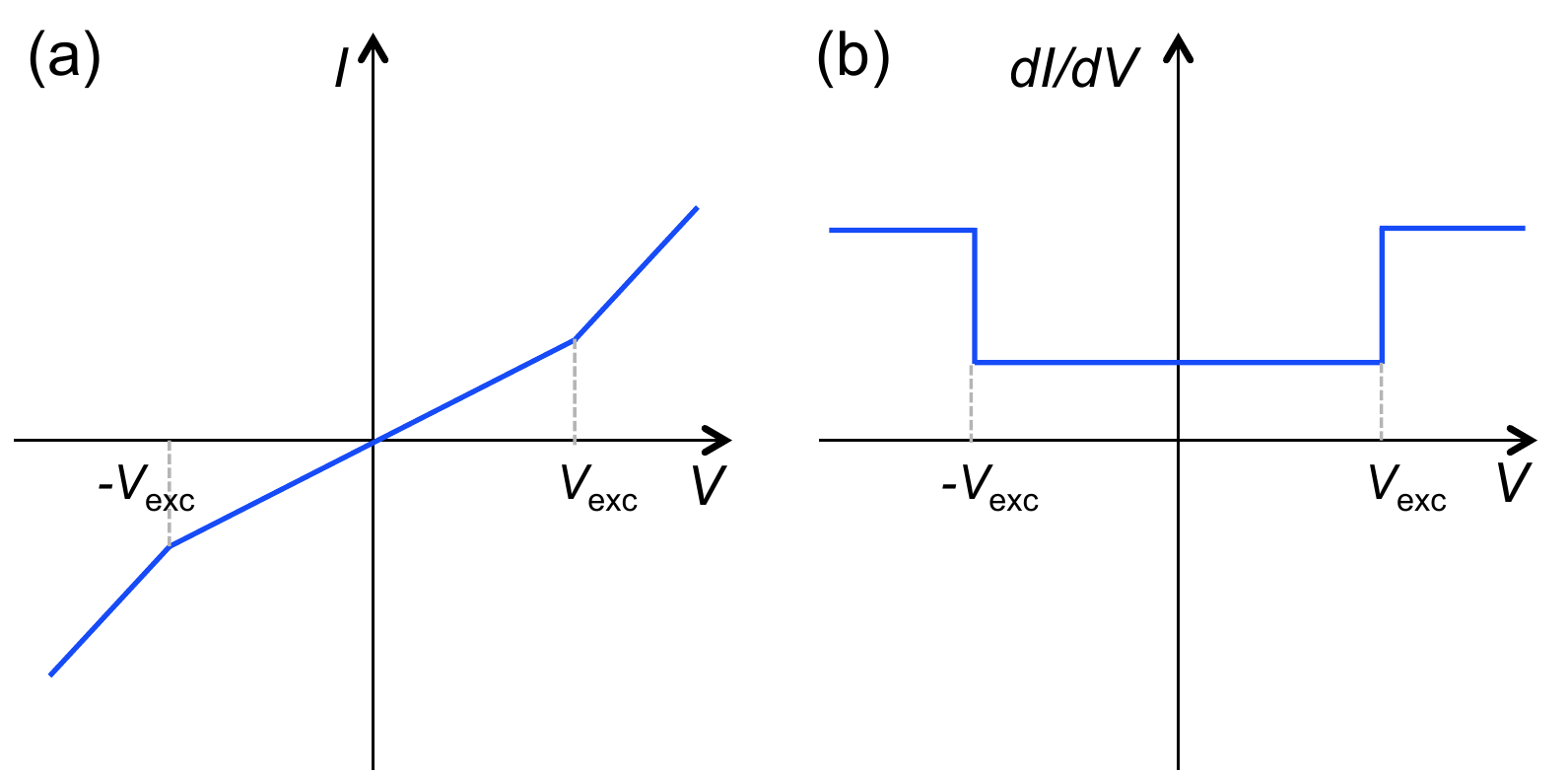}
\caption{\label{IETS}
(a)~Current versus voltage spectrum with an additional tunnel
channel at a threshold voltage $|V_\mathrm{exc}|$ due to an inelastic
excitation. (b)~Inelastic excitations are typically studied measuring the
differential conductance and then appear as symmetric steps at $-V_\mathrm{exc}$
and $V_\mathrm{exc}$ around the Fermi-energy.  } 
\end{figure}

Figure~\ref{IETS} $(a)$ and $(b)$ shows typical IETS measurements. When the
bias matches an excitation energy, $V_\mathrm{exc}=E_\mathrm{exc}/e$, the tunneling
electron can yield part of its energy and end up in a different
state. The effective number of final states for tunneling suddenly
increases at the threshold $V_\mathrm{exc}=E_\mathrm{exc}/e$. 
As a consequence,
the tunneling current changes its slope, Fig.~\ref{IETS} $(a)$,
which is more clearly seen in its derivative, $dI/dV$,
or differential conductance, Fig.~\ref{IETS} $(b)$.
The steep increase at threshold and the electron-hole symmetry
of the IETS signal are the hallmarks that the spectral
features in the differential conductance correspond to an excitation.

The above properties are common to any kind of excitation that can
be induced by tunneling electrons. Spin can flip under a tunneling
electron, giving rise to magnetic excitations that 
can be detected in the
same way~\cite{heinrich_2004}. This is of great value in the study
of spin chains because 
it furnishes a characterization of the spin chain.
Typical spin-flip excitations are in the meV range, where they can
coexist with acoustic phonons that are difficult to excite by tunneling
electrons~\cite{Karina}. Contrary to phonons, spin-flip excitations
are very easy to excite. A simple picture relates the change in
conductance over the excitation threshold with the fraction of tunneling
electrons that actually induce the excitation~\cite{Lorente_2005}.
While vibrational excitation yields excitations in the range of 10\%,
magnetic excitations easily exceed 100\%~\cite{Lorente_2009}.


Recently, it has been shown that IETS can also detect orbital
excitations. In this case, the symmetric signature of excitations
in the IETS is also lifted because it depends on the occupancy of the
orbitals~\cite{Kugel_2018}.

\textit{Spin-dependent tunnel processes.---}
\nico{When a magnetic tip is used, the tunneling current can be spin-polarized
(SP). This has implications both
for the elastic as well as the inelastic contribution to the tunnel
current~\cite{Bode_2003,Wiesendanger_2009,Loth_2010,Loth_2010b}. For a static magnetization
of a sample the spin polarization of the tunnel current and the
differential conductance scales with the projection of sample
onto tip magnetization, i.e.\ a tip magnetized along its axis is
sensitive to the out-of-plane component of the sample magnetization,
whereas a tip magnetization parallel to the surface plane detects
in-plane magnetization components of the sample. These so called
spin-resolved STM (SP-STM) and spin-resolved STS (SP-STS) modes allow
access to the spatially resolved magnetic properties of magnetic atoms,
nano-structures, or surfaces down to the atomic scale~\cite{Bode_2003,
Wiesendanger_2009,Wiebe_2011,Bergmann_2014}. 

As we saw for IETS, when the tunneling electron can
induce a spin flip excitation in the tunnel junction the spin polarization
of a tunnel current leads to a preferred direction of excitation, i.e.\
the minority spin channel of the tip can flip a spin in one direction and
the majority electrons flip it in the opposite direction. This leads to
the above lifting of electron-hole symmetry by having
different amplitudes of the inelastic excitation steps at
positive and negative bias in the differential
conductance. The asymmetry scales with the spin-polarization of
the tunnel current for low tunneling rates.   
An additional source of bias asymmetry comes from the
spin-conserving potential scattering that leads
to interference with the spin-flip contribution.
At higher tunneling rates,
spin pumping can occur, because multiple subsequent excitations
survive before de-excitation. This drives the
system out of equilibrium with
sizable bias asymmetries~\cite{Loth_2010,Loth_2010b}.
}

\begin{figure}
\includegraphics[width=1.0\columnwidth]{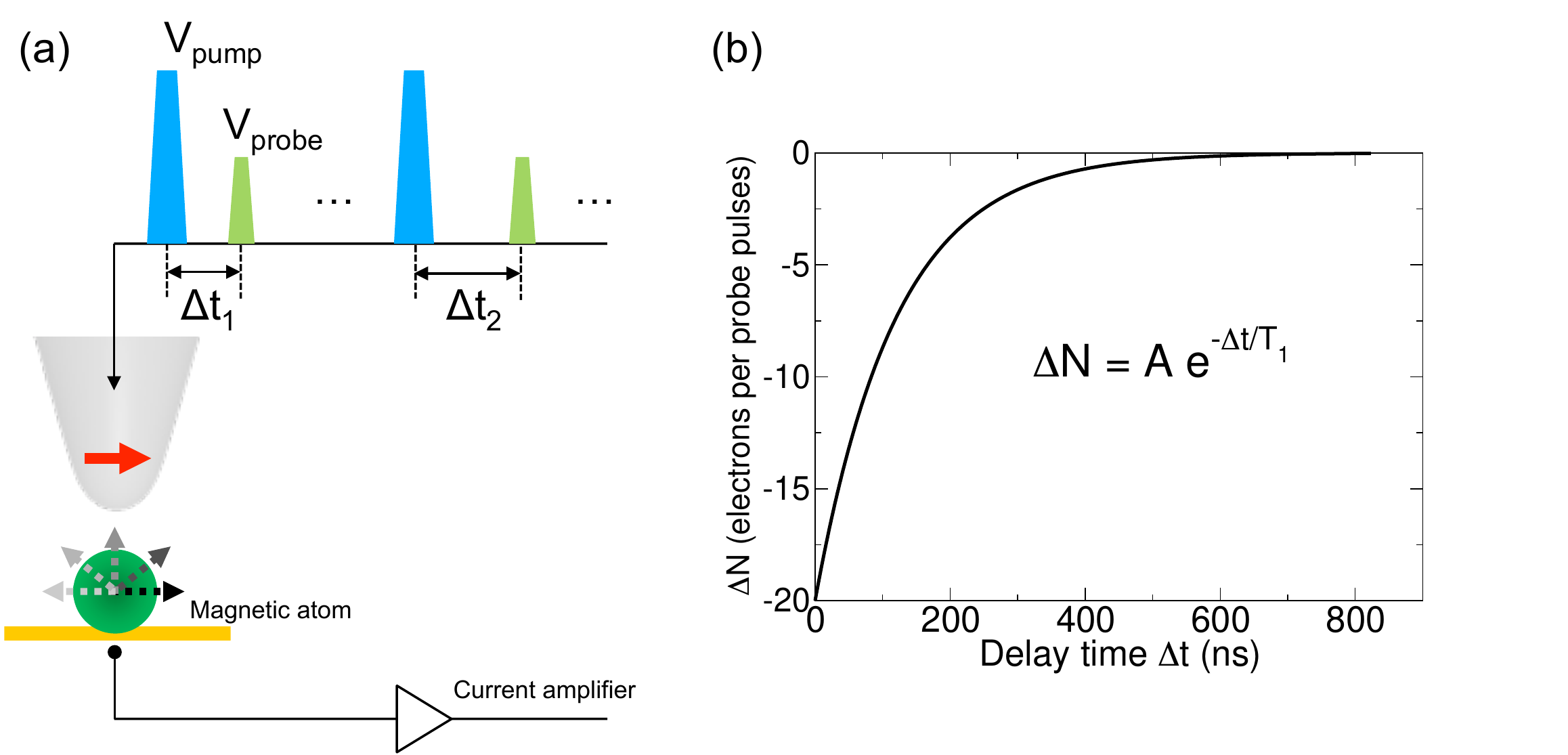}
\caption{\label{pump_probe}
Simplified diagram of the electronic pump-probe technique (adapted
from ~\cite{Lot2010}).  (a) First, the pump pulse excites the
spin states of the adsorbed magnetic atom and second,  the probe
pulse at certain time delay ($\Delta$t$_1$) detects the status of the
spin by spin polarized tunnelling. By varying the time delay ($\Delta$t)
and sending repeated sets of pump-probe pulses, (b) the conductance
as a function of a time is obtained, which gives  information of the spin
relaxation time (T$_1$). In the present example the initial 
number of collected electrons is $-20$ and the relaxation time
is 120 ns.
}
\end{figure}

\textit{Pump-probe techniques.---}
\nico{The dynamical evolution of spin excitations can be observed by using the previous
pumping process. This has grown to become an STM-based 
electronic pump-probe technique.  One of the
first applications was to measure the spin
relaxation time of a  Fe-Cu coupled dimer on Cu$_2$N surface~\cite{Lot2010}.

The technique uses a series of electronic pump and probe pulses that
are generated and sent to the STM (see Fig.~\ref{pump_probe} $(a)$).
Once the electronic pump-pulse is sent to the adsorbates on the surface,
the spins of the system excite and relax over time.  The voltage of the pump
pulse has to be larger than the excitation energy to excite the spin
from the ground state to an excited state by inelastic scattering
of tunnelling electrons.  A probe pulse of smaller voltage is sent
to measure the state of the spin. This is achieved by magnetically
polarizing the STM tip.  By sending the probe pulse at different time
delays ($\Delta$t), information on the
time dependence of the population of the levels can be obtained by measuring the evolution of conductance 
with $\Delta$t (Fig.~\ref{pump_probe} (b)). The conductance
behaves exponential with $\Delta$t, a characteristic time constant given by the spin
relaxation time (T$_1$).  The working principle of this technique is 
tunnelling magneto-resistance.  After the spin is excited by a pump pulse,
the spin relaxes and goes to the ground state. Using a spin-polarised
tip, depending on the orientation of the adsorbed spin at certain
time delay, the conductance will change with the characteristic T$_1$
time constant.

Using this technique, the spin relaxation time of Fe spin chains
have been measured as a magnetic tip was being approached~\cite{Yan_2015}.
The exchange field of the tip changed the state mixing of
the spin chain, and this had a measurable effect on the
lifetime of the spin chain excitations.}

\subsection{Preparation of chains on surfaces}

There are two possibilities to obtain well defined
chains on surfaces: self-organization or atom
manipulation. A spontaneous formation of spin chains can be
realized on uniaxial surfaces, taking advantage of surface
thermodynamics~\cite{Himpsel,Gambardella_2000,Gambardella_2002}. In
particular metallic substrates allow for the required atom diffusion
for self-organized growth and it was shown that tens of nanometer-long
one-dimensional spin chains can be reproducibly achieved. Atom
manipulation with the tip of an STM was first demonstrated for Xe
atoms on a Ni(110) surface~\cite{eigler_1990}. In both lateral and
vertical manipulation mode the force between the tip and a single atom
adsorbed on a surface is exploited to reproducibly displace the single
atom~\cite{Bartels_1997}, making it possible to build nano-structures
atom by atom~\cite{morgenstern_2013,Lorente_2005}. 
Typically the potential landscape for diffusion on metal
surfaces is smooth enough to allow for lateral manipulation of adatoms,
whereas surfaces that form covalent bonds with the adatom such as 
a Cu$_2$N layer grown on Cu(100), that has been the substrate for
various magnetic chains as discussed below, requires the pick-up and
drop-off of single atoms with the tip, i.e.\ vertical manipulation.

\subsection{Spin chains and their holding substrates}

In this Colloquium, we focus on atomic spin chains on a surface. In
this scenario, the influences of the substrate on the (magnetic)
properties of the spin chain become an important consideration, and
one of the main factors is the coupling strength to the substrates
electron bath. When a spin chain is only weakly coupled to an electron
bath, as for lightly-doped semiconductor substrates, the low electron
concentration impedes electronic excitations of the low-energy magnetic
states of a spin chain. Only phonons are available for damping and
they are not very efficient because $(i)$ they need a large spin-orbit
interaction to couple spin and atomic-position degrees of freedom
and $(ii)$ the number of available phonons is very limited at low
temperatures. However, in lightly-doped semiconductors spin excitations
are difficult to detect experimentally, due to practicalities related
to detecting changes in conductance when the applied bias is large
enough to overcome the electronic band-gap~\cite{Alex_Nature}. To
circumvent this, a metal substrate passivated with a semiconducting
or insulating film can be used as substrate for magnetic chains. In
such a sample, the passivation of the metal substrate reduces the
coupling with the electron bath, hence increasing the lifetime of the
intrinsic spin-chain excitations while permitting good conductance
conditions to resolve the electronic current from the STM tip. A
particularly successful substrate for the construction of extended
spin structures has been the case of a monolayer of Cu$_2$N grown on Cu
(100)~\cite{Hir2006,Loth_2010,spi2014,Hir2007,Otte_2008,Otte_2015,Choi_2016,Choi_2016b}
(see Sec.~\ref{Decoupled}).

On the other hand, in order to strongly couple the magnetic chain
to an electron bath, the chain atoms can be adsorbed directly
to a metal substrate, which efficiently couples their orbitals
to the delocalized electrons of the substrate. In this case, any
excitation of the system is damped relatively quickly due to the
enhanced coupling. The damping occurs via efficient electron-hole
excitations, which dominate over other de-excitation channels even at
very low temperatures. When an atomic system of spins is in contact
with a metal surface, the magnetic spectra become broadened by the
above mechanism. This leads to broad features in the differential
conductance spectra which, nevertheless, can be still detected with
the STM~\cite{balashov_2009,schuh_2010,Alex_Cu,Alex_Ag,Alex_Pt}. At
the same time the strong hybridization with the substrate can lead to
induced magnetic moments in the substrate atoms, and often the magnetic
properties can be understood within the classical limit. In addition
to the investigation of dense and dilute chains on such normal metal
surfaces, there has recently been increased interest in the properties
of magnetic chains on superconductors (see Sec.~\ref{Shiba}).

%% file: SpinChainsDecoupled/passivated.tex
\section{Spin chains decoupled from the substrate's electron bath}
\label{Decoupled}

\subsection{Passivated metal substrates}
\label{passivated}

The first demonstration of spin excitations on a passivated metal substrate was performed by Heinrich and coworkers in 2004~\cite{heinrich_2004}.
The authors used Al$_2$O$_3$ islands grown on NiAl to deposit
a small number of Mn atoms where they performed conductance
measurements as a function of bias at low temperature (0.6 K)
and with magnetic fields as high as 7 T. Other, more recent experiments involve single atoms and small multi-atom structures on MgO on Ag(100)~\cite{Rau_2014,Baumann_2015,Baumann_2015b,Natterer_2017}, providing the opportunity to tune the coupling strength to the substrate by varying the number of MgO layers. A particularly successful substrate for the construction of extended spin structures has been the case of a monolayer of Cu$_2$N grown on Cu (100)~\cite{Hir2006,Loth_2010,spi2014,Hir2007,Otte_2008,Otte_2015,Choi_2016,Choi_2016b}, which will be the main focus of this section.

Bulk copper-nitride is an insulator with a
gap of above 4~eV. A single atomic
layer does not form a complete insulator and
only partially decouples the spin from the copper metal
substrate. These conditions turn out to be ideal for inelastic tunnelling spectroscopy (IETS) experiments.
In addition to acting as a decoupling layer, the Cu$_2$N surface forms a good template grid for assembling spin chains. The N atoms are bonded covalently to the Cu atoms, making the Cu$_2$N layer essentially a two-dimensional molecule with square symmetry~\cite{Hir2007}. When a transition metal atom, such as Co, Fe or Mn, is positioned on top of the layer, it is incorporated into that molecule. As such, manipulation of adatoms on top of Cu$_2$N can be seen as the construction of a two-dimensional magnetic molecule with spin centers at predeterminable locations.

Density functional theory (DFT) calculations~\cite{Choi_2016,Rudenko_2009,urdaniz_2012} show that
the Cu$_2$N monolayer is profoundly modified when a magnetic atom is placed directly over a Cu atom -- the typical binding site for transition metal atoms. The Cu atom
underneath the magnetic atom is pushed into the substrate
while the two neighbouring N atoms are pulled upwards into
the chain. As a consequence, we can view a spin chain built on Cu$_2$N
as an ensemble of alternating transition metal (TM) atoms and N atoms. 

The crystal field due to the nitrogen network can provide an anisotropic environment with clear preferred magnetization axes for the spins~\cite{Hir2007}. For the magnetocrystalline anisotropy encountered on the Cu$_2$N surface, typically the second order form of Eq.~(\ref{MAE}) involving a uniaxial parameter $D$ and a transverse term $E$ suffices. A study by Bryant et al. \cite{Bryant_2013} showed that these phenomenological parameters can be understood readily in terms of the angle between the two nitrogen bonds pointing away from the magnetic atom. The closer this angle is to 180$^\circ$, the larger the ratio $D/E$. The exact geometry of atoms incorporated into the network will be discussed further down.

The molecular nitrogen network is also responsible, at least in part, for mediating spin-spin coupling over distances spanning several unit cells~\cite{Otte_2008}. By placing pairs of magnetic atoms near each other on the network, depending on their relative positioning different coupling signs and strengths are found with values of the Heisenberg parameter $J$ ranging from $J\sim+2$~meV (antiferromagnetic) to $J\sim-1$~meV (ferromagnetic)~\cite{spi2015}. While the exact physical mechanism underlying the spin-spin coupling remains to be studied further, it is believed to be a combination of superexchange mediated by the nitrogen network and Ruderman-Kittel-Kasuya-Yosida (RKKY) coupling~\cite{Yosida} mediated through the metal underneath. In general, it is found that the coupling strength decreases rapidly with the number of nitrogen bonds separating the atoms.

\begin{figure}
\includegraphics[width=1.0\columnwidth]{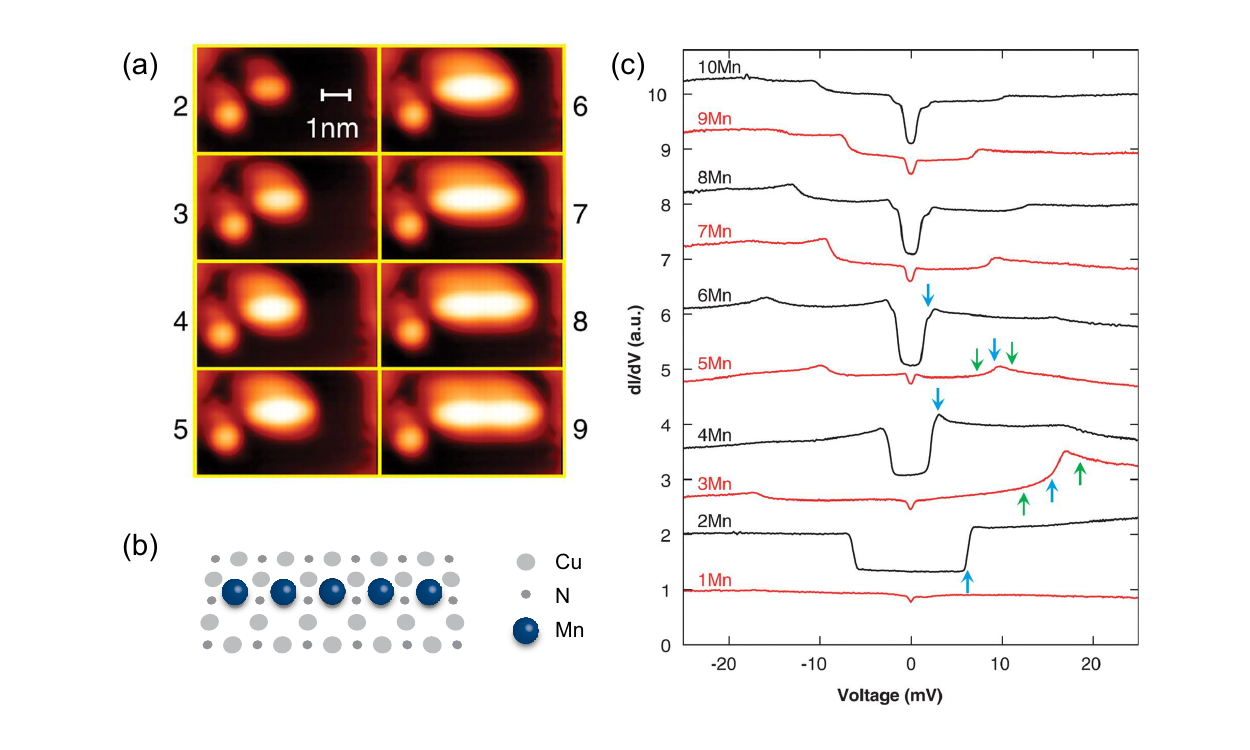}
\caption{\label{fig:Mn_chain}
Constant current images, $(a)$, of the created Mn$_n$ chains with $n=2\cdots9$ on Cu$_2$N/Cu(100).
Differential conductance over atomically manipulated Mn$_n$ chains with $n=1\cdots10$, $(b)$.  
Depending on the number of Mn atoms, the behavior is different. For odd number, it shows a small bias feature with the spin changing excitation energy steps 
while for even number,
it only gives the spin excitation energy steps. The lowest spin changing excitations are marked in blue arrow.
Figure adapted from \cite{Hir2006}.}
\end{figure}

The first spin chains on Cu$_2$N were built by Hirjibehedin
and coworkers. They showed that Mn$_n$ chains could be
built with $n=2\cdots10$ by using a vertical atom manipulation
technique~\cite{Hir2006} (Fig.~\ref{fig:Mn_chain}(a)). The atoms were placed one unit cell apart, being separated by a single N atom (Fig.~\ref{fig:Mn_chain}(b)). IETS showed clear and distinct behaviour
depending on the parity of $n$. For chains with an even number of Mn
atoms, clear excitation thresholds appeared at several meV, that reduced in energy
as the number of Mn atoms increased. Odd-numbered chains, on the other hand,
displayed a small-bias featured reminiscent of the small magnetic
anisotropy of a single Mn atom (Fig.~\ref{fig:Mn_chain}(c)). These spectra were readily
explained in terms of an isotropic Heisenberg Hamiltonian with antiferromagnetic coupling between the spins in the chains. In contrast to other spin chains discussed below, in the case of these close-spaced Mn chains spectroscopy was found to be the same regardless of the position of the tip on the chain. As such, the chains could be viewed as a single magnetic entity, using the giant spin approximation. For even-numbered chains, having zero total spin in the ground state, the observed excitation corresponds to a singlet-triplet excitation. Odd-numbered chains, on the other hand, have a net spin of 5/2. Similar to the single Mn atom, their spectra display only a small-bias dip. Indeed, the exchange interaction obtained in this way matched the computed exchange interactions for the same system~\cite{Rudenko_2009,urdaniz_2012}. 

Follow-up work focused predominantly on Fe chains. Loth \textit{et al.}~\cite{Loth_2012}
showed that for chains with an interatomic spacing of two unit cells (which can essentially be seen as repetitions of a Fe-N-Cu-N unit cell), the magnetic ordering was still antiferromagnetic. But in contrast to the earlier closed-spaced Mn chains, spectroscopy performed on each of the atoms in the double-spaced Fe chains showed different excitation intensity, justifying a description in terms of weakly coupled local magnetic moments. Spin-polarized measurements indicated that the two lowest-energy states of the Fe spin chains are N\'eel states that correspond to a classical arrangement of opposing spins.
Under influence of either tunneling current or temperature, switching between the two possible
 N\'eel states could be induced. The study of the efficiency of the switching as a function of applied
bias permitted the authors to determine a threshold and hence
identify an indirect mechanism for switching. An impinging electron would excite
the spin chain into an excited state followed by a decay into
the \textit{other} N\'eel state. Calculations proved that collective
excitations were at play~\cite{Gauyacq_2013}. The threshold was determined
by the excitation of the lowest-energy spin wave of the chain. As the energy
of the tunneling electron increased, more excitations of the chain could be excited
improving the switching mechanism to the point that a 50\%-50\% de-excitation probability into
either of the two N\'eel states was reached~\cite{Gauyacq_2013}.

Advanced measurements on these spin chains showed that the combination of IETS and the sudden variation in spin-polarized current due to the change in state population near the excitation threshold could lead to peculiar spectroscopic features including negative differential conductance~\cite{Rolf-Pissarczyk_2017}. In addition, it was shown that the chains could be flipped as well due to the effect of exchange bias with the magnetized STM tip, provided that the tip was brought in sufficiently close proximity to the structure~\cite{Yan_2015}.

\begin{figure}
\includegraphics[width=1.0\columnwidth]{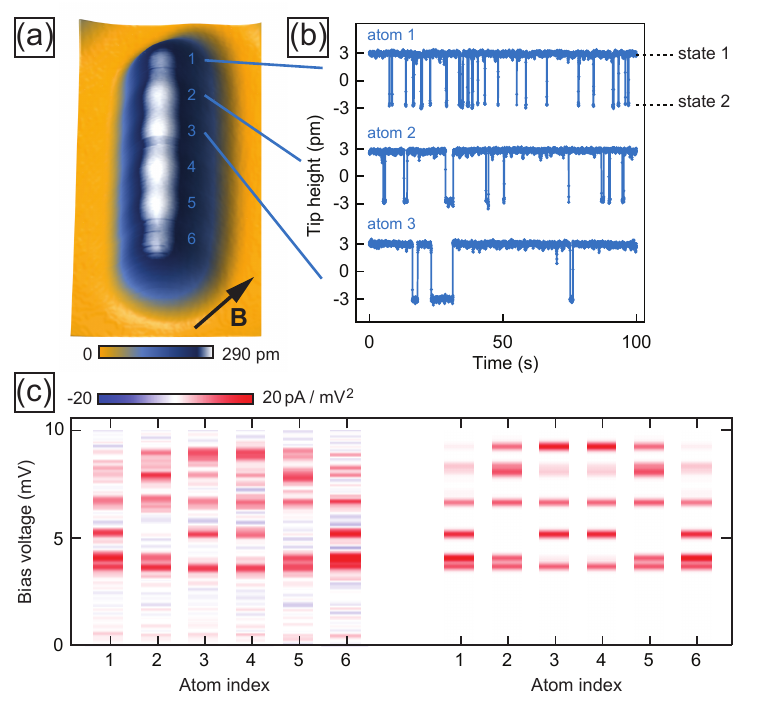}
\caption{\label{fig:spinwaves}
Detection of spin waves in a ferromagnetic chain. (a)~STM topography of a ferromagnetic 6-atom Fe chain on Cu$_2$N/Cu(100). (b)~Telegraph noise measured using spin-polarized STM on the first three atoms of the chain. Switching is observed between two metastable states. The switching rate decreases as the tip is moved towards the center of the chain. (c)~Left: IETS spectra taken on each of the atoms in the chain. Spin wave states are observed with recognisable nodal structure at $\sim3.5$~mV, $\sim4.0$~mV and $\sim5.5$~mV. Right: corresponding theory obtained from diagonalization of the spin Hamiltonian. Figure reproduced from (\cite{spi2014}).}
\end{figure}

An experimental study focused on the collective excitations that
are populated during the switching process was provided by Spinelli
\textit{et al.}~\cite{spi2014}. Here, the authors studied chains of Fe
atoms that, due to a different orientation of the chains on Cu$_2$N, were
coupled ferromagnetically (Fig. \ref{fig:spinwaves}(a)). The resulting
two metastable states were states where all the spins were parallel to
each other and pointing in one of the two opposing directions along the
easy axis. Also here, telegraphic switching between the two metastable
states was observed (Fig. \ref{fig:spinwaves}(b)). In particular, the
switching was found to be current-induced and dependent on the location
of the STM tip over the chain. IETS measurements performed on each of the atoms in the chain revealed that
the lowest energy excitations were of spin wave nature: a clear nodal
structure was observed, with the number of nodes increasing for higher
energy modes (Fig. \ref{fig:spinwaves}(c)). Rate equation calculations
indicated that the lowest energy transitions from one metastable state to
the other passed through a sequence of these spin waves states, followed
by domain wall states sweeping the domain from one end of the chain to
the other~\cite{spi2014}.

Spin chains made of Co atoms have shown a very different behavior~\cite{Otte_2015}.
Intriguingly, clear IETS measurements could be performed on \textit{only} the edges of
the spin chains, while no signal was recorded over the bulk sites. The explanation of this peculiar behaviour lies in
the actual geometry of the chain: the edge atoms have a finite N-Co-N angle, whereas for the atoms away from the edges the N-Co bonds were almost collinear. This leads to an electronic structure where there is no overlap between the tip apex and the d-orbitals of the bulk Co atoms. As a result, interaction of the tunneling electrons with the local spins of the chain is avoided, preventing spin excitations from occurring~\cite{Otte_2015}.

\begin{figure}
\includegraphics[width=1.0\columnwidth]{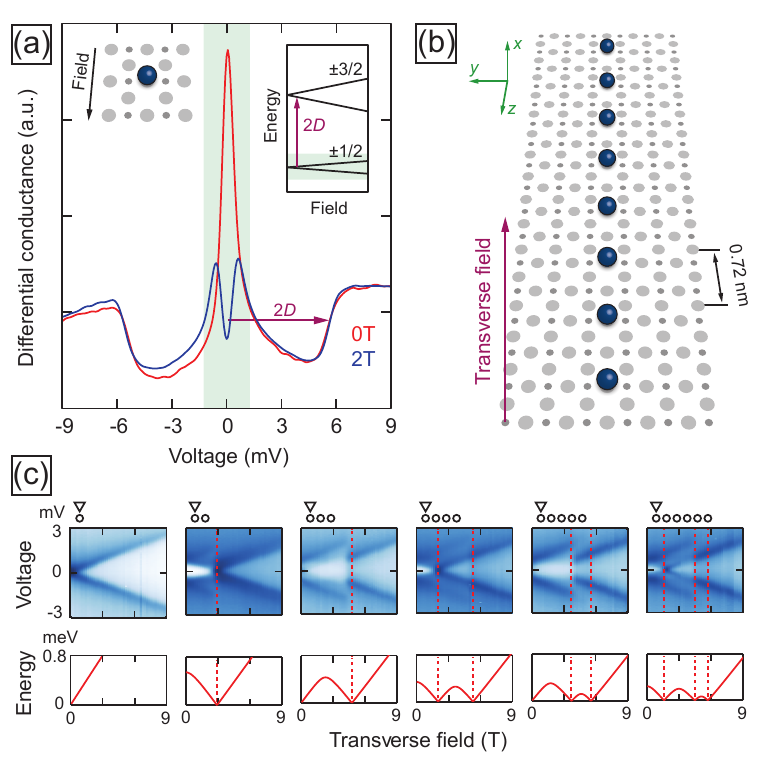}
\caption{\label{fig:xxz}
Using an atomic spin chain for quantum simulation. (a)~IETS spectrum taken on a single Co atom on Cu$_2$N/Cu(100), indicating a split between the $\pm 1/2$ and $\pm 3/2$ doublets. (b)~Atomic design of an XXZ Heisenberg chain in transverse field using Co atoms. (c)~Top: IETS performed on the first atom of chains up to length 6 for transverse magnetic fields up to 9~T. Transitions in the ground state are observed leading up to the critical point near 6~T, which coincide with theoretically predicted ground state changes (bottom). Figure reproduced from (\cite{tos2016}).}
\end{figure}

Recently, Co chains of a different configuration were shown to be a useful platform for making experimental realisations of model spin Hamiltonians~\cite{tos2016}. While Co atoms in principle have a spin magnitude $S=\frac{3}{2}$, here the authors demonstrate that an effective spin-$\frac{1}{2}$ chain can be engineered by making use of magneto-crystalline anisotropy. Co atoms  on Cu$_2$N are found to experience hard axis anisotropy, as a result of which the $m=\pm\frac{1}{2}$ Kramers doublet is split off approximately 6~meV below the $m=\pm\frac{3}{2}$ doublet (Fig. \ref{fig:xxz}(a)). By designing the antiferromagnetically coupled chain such that the coupling strength $J$ between the spins is much smaller than 6~meV, an effective spin-$\frac{1}{2}$ chain with anisotropic XXZ coupling is formed~\cite{tos2016} (Fig. \ref{fig:xxz}(b)). The model XXZ Heisenberg Hamiltonian is known for a critical point at a certain value of the transverse magnetic field, beyond which the chain becomes paramagnetic. Before reaching this critical point, the system is characterised by a ground state doublet which is topologically separated by an excitation gap from the continuum of states~\cite{Dmitriev_2002}. Local spectroscopy measurements on the Co chains as a function of the transverse magnetic field revealed these two states and their interplay in the region leading up to the critical point (Fig. \ref{fig:xxz}(c)).

%% file: SpinChainsDecoupled/kondo.tex
\subsection{The impurity problem and its extension to spin chains}
\label{Kondo}


\nico{
An interesting twist to the above results comes when heterogeneous spin chains
are used.  Indeed, if we can  
 consider a spin chain as a single magnetic object in
contact with a reservoir of electrons (the substrate), a Kondo
effect due to the collective
behavior of the spin chain can take place.
The Kondo effect \cite{Hewson} is due to the
electronic correlations caused by spin-flip scattering
off a  magnetic impurity.
The magnetic impurity has to present a two-fold degenerate
ground state in the absence of spin flips that can be switched
one into the other by  zero-energy
spin flips~\cite{Choi_2018b}. In order to achieve this
in a spin chain, all spins need to be strongly entangled.
}

\nico{
In the previous section, we showed the case of 
Mn chains~\cite{Hir2006,Choi_2016}.  Chains
with an odd number of Mn
atoms have a 5/2 ground states that cannot be connected via $\Delta
S_z=\pm 1$ spin flip. Hence, no Kondo effect takes place.  
Even-numbered antiferromagnetic chains are
singlets so they have no degeneracy. As a consequence, no Kondo effect is revealed in the dI/dV spectra
of these chains. Other antiferromagnetically coupled chains such as
Fe$_n$~\cite{Loth_2012,spi2015} and Co$_n$~\cite{Otte_2015} show no
degenerate ground state either preventing the formation of Kondo correlations.
It seems difficult to have a spin chain with a
degenerate ground state that can be connected via spin flips.
The solution was found in \cite{Choi_2017c} by building
heterogeneous chains with two ingredients: $(i)$ 
a two-fold degenerate spin ground state in the absence of spin flips, and $(ii)$ strong entanglement
such that a single spin-flip from a substrate electron can reverse the
ground state.  
}

\begin{figure}[t]
\centering  \includegraphics[width=0.5\textwidth,clip=]{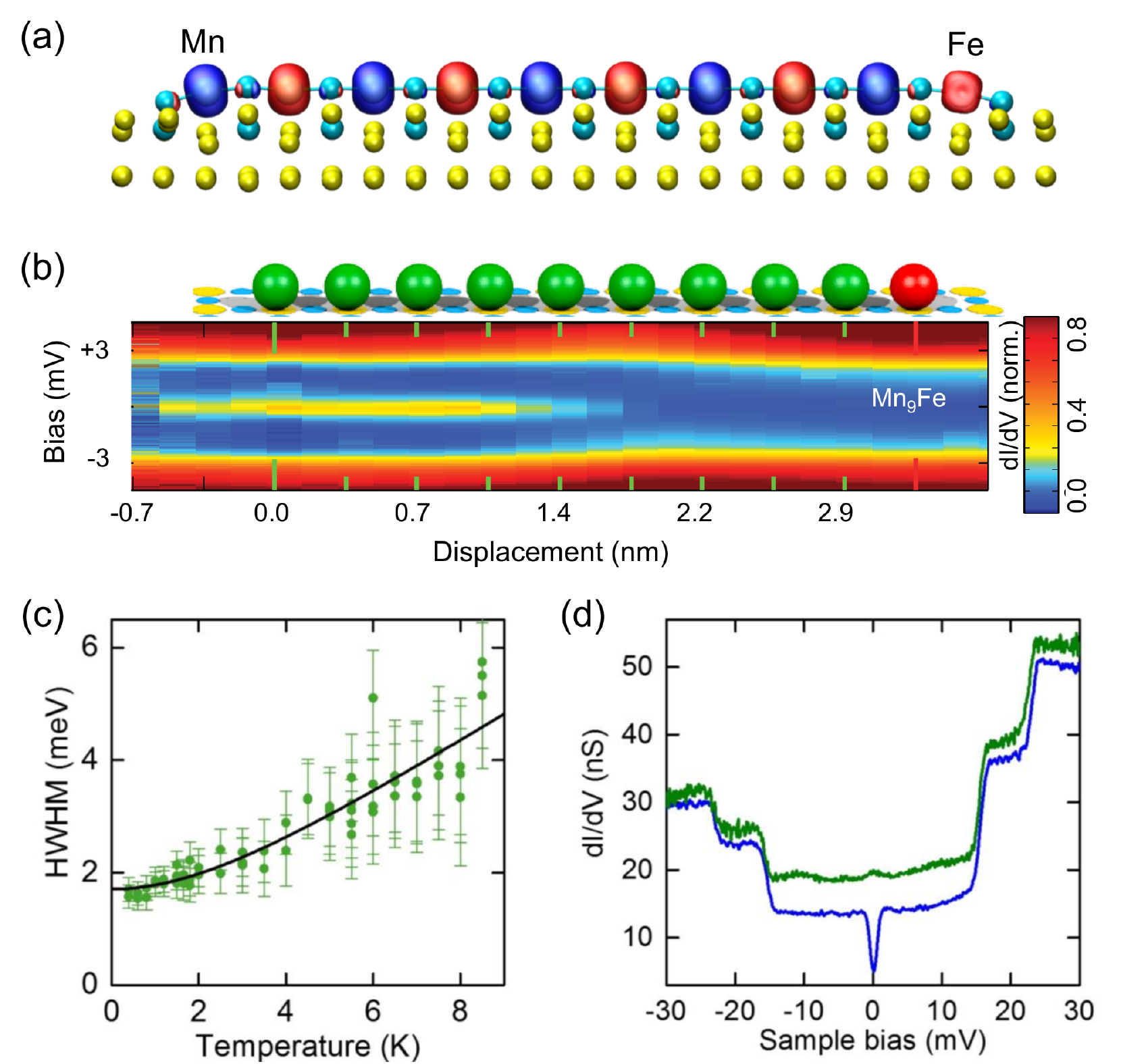}\\
 \caption{
Heterogeneous FeMn$_n$ spin chains ($n=1-9$), on Cu$_2$N / Cu (100).
(a) Density difference between spin-up (blue) and spin-down (red) states over a Mn$_9$Fe chain. 
Copper atoms (yellow) and Nitrogen atoms (cyan). 
(b) Differential conductance (dI/dV) map along Mn$_9$Fe chain. 
dI/dV signals plotted as a function of sample bias (V) and displacement (nm).
Mn(green ball) and Fe(red) atoms are visualized to show where they sit. 
$(c)$ Half width at half maximum (HWHM) (meV) of the zero-bias anomaly for MnFe dimer plotted as a function of temperature. 
The black line is a fit to the Kondo peak as a function of temperature. 
$(d)$ The differential conductance as a function of applied bias for two different magnetic fields in MnFe dimer.
The green curve corresponds to no magnetic field and the blue one to a $B=5T$ field applied
perpendicular to the surface. The measuring temperature is 0.5 K. The green
line is vertically shifted by 5 nS for clarity. Figure adapted with permission from~\cite{Choi_2017c}.
Copyright (2017) American Chemical Society. 
}
\label{FeMn}
\end{figure}

In \cite{Choi_2017c} the authors built Mn$_n$ chains where they
added a terminal Fe atom. The newly created FeMn$_n$ chains were
in principle $S=1/2$ systems for an odd number of Mn atoms (odd $n$)
 since all atoms couple antiferromagnetically
along the nitrogen rows of the Cu$_2$N / Cu (100) substrate.
The same could also be said of Fe$_n$Mn chains, since
again, for odd $n$ the sum of antiferromagnetically aligned  magnetic moments
leads to $1/2$. However, the experiment
showed that these two systems behave very differently. In the case of $n=3$,
the FeMn$_3$ chain displayed a zero-bias anomaly that was shown to be
a Kondo peak while the Fe$_3$Mn chain displayed no Kondo peak~\cite{Choi_2017c}.
The authors realized that besides the exchange interaction
controlling the spin-spin coupling along the chain, the magnetic anisotropy
of each atomic spin was important. Indeed, Fe presents a large axial magnetic
anisotropy as compared to Mn. Assuming
similar exchange couplings, the Fe-rich chains will tend
to align their spins more than the Mn-rich chains. Thus, a Fe$_n$Mn chain will have
a larger number of spins that prefer to align along the atomic spin axis,
leading to an Ising-like spin system, and hence to a system with  reduced entanglement.
Flipping the full spin of the Fe$_n$Mn chain via a substrate spin flip
becomes very difficult. However, the magnetic
anisotropy of FeMn$_n$ is very reduced and the ground state strongly resembles the
one of an antiferromagnetic Heisenberg chain, strongly entangled. As
a consequence a single spin flip from the substrate has a larger
probability of flipping the full spin of the chain leading to the
Kondo phenomenon.

\nico{
Figure~\ref{FeMn} summarizes the behavior of  
 FeMn$_n$ spin chains ($n=1-9$) on Cu$_2$N / Cu (100).
Similarly to the Mn$_n$ spin chain, the atomic structure
of the chain includes strong relaxation of the surface
as shown in the results of DFT calculations plotted in Fig.~\ref{FeMn} $(a)$.
There the incorporation of N atoms into the chain is evident as
well as the reorganization of the nearest Cu atoms.
Despite the chain being mostly a Mn$_n$ spin chain, the addition
of an extra Fe changes the spectral features. In Fig.~\ref{FeMn} $(b)$
a clear Kondo feature is localized on the edge of the chain
that \textit{does not} contain the Fe atom despite
the fact that without Fe, there is no Kondo peak. 
This behavior can only be explained if indeed the chain
is reacting like a single object allowed by the
entanglement of spins. That the peak at zero bias
is indeed a Kondo peak is shown in Figure~\ref{FeMn} $(c)$
where the behavior of the zero-bias peak with temperature,
follows the trend of a Kondo peak. 
The data are fitted by a function, $\Gamma_K (T)$, that takes into
account the thermal broadening of the resonance and of the tip's
Fermi function~\cite{Nagaoka_2002,Ternes_2009}:
\begin{equation}
\Gamma_K (T)=\frac{1}{2} \sqrt{(2 \Gamma_K^0)^2+(3.5 k_B T)^2+ (\alpha k_B T)^2}.
\label{crommie}
\end{equation}
The coefficient $\alpha$ reflects how close to a Fermi-liquid solution
the Kondo system is~\cite{Nagaoka_2002,Ternes_2009,Ternes_2015}. The fit
of Fig.~\ref{FeMn} $(c)$ reveals a $T=0 K$ Kondo width, $\Gamma_K^0$,
of 1.68 meV and a $\alpha$  value of 11.1. These
values agree with the behavior of a Kondo system. Moreover, the large value for $\alpha$
points at the behavior of a Kondo system
in the weak coupling regime~\cite{Ternes_2013}.
Further evidence can be found in Fig.~\ref{FeMn}$(d)$
where the effect of the magnetic field splits the Kondo peak as expected.
When a magnetic field of 5 Tesla is  perpendicular to the sample, the $S=1/2$-like ground
state splits and the elastic spin flips giving rise to the Kondo peak cannot
be produced anymore. The Kondo peak disappears and instead a clear
inelastic spin-flip signal develops as can be seen in the blue line of the graph.
}

%% file: SpinChainsDecoupled/semiconductors.tex
\subsection{Semiconductor substrates}
\label{SemiSub}

Semiconducting substrates principally offer a huge flexibility for the tuning of the properties of spin chains. The substrate electron density, and thereby the coupling of the chain spins to the electron bath, can be largely adjusted by the doping of the used semiconductor materials. Thereby, it is essentially possible to \nico{investigate the continuous transition} from the decoupled spin case of the passivated surfaces to the strongly coupled case of the metallic substrate (Section~\ref{Coupled}). The (110) surfaces of narrow gap III-V semiconductors, e.g.\ InAs and InSb, additionally feature the possibility to change the dimensionality of the electron bath between 3-D, 2-D, 1-D, and even 0-D, by surface doping, polar step edges or using the STM tip induced quantum dot~\cite{Morgenstern_2003, Meyer_2003, Wiebe_2003, Hashimoto_2008}. Despite all these advantages, studies of spin chains on semiconductors are yet relatively sparse due to preparation and measurement difficulties.

For (110) surfaces, STM-tip based manipulation is rather uncontrolled because they form strong covalent bonds with metal adsorbates~\cite{Kitchen_2006, Kitchen_thesis, Richardella_2009, Gohlke_2013}. F\"olsch and co-workers~\cite{Folsch_2009,Folsch_2014, Folsch_2015, Yang_2012} have succeeded in the creation of individual chains of metal atoms on the (111) surface of MBE grown InAs by STM-based atom manipulation. STS of their electronic properties demonstrates the fascinating possibility to control the chain's and substrate's electronic properties down to the single atom limit. However, the spin properties of such chains have not yet been studied.

Self-organized growth of metal chains is limited to few substrates \cite{matsui_2007, Snijders_2010, Houselt_2013}. STS investigations of the electronic properties of individual gold chains grown by self-assembly on stepped~\cite{Crain_2005} and flat~\cite{Yeom_2015} Si substrates have been performed. The authors conclusively show the 1-D character of confined electron states in gold chains. Although the states in such chains have been predicted to be spin polarized~\cite{Erwin_2010} investigations using SP-STS or IETS are, so far, lacking. Matsui and collaborators~\cite{matsui_2007} describe room temperature deposition of Fe on the (110) surface of InAs prepared by cleavage in UHV. This procedure results in the self assembly of sparse short chains with a maximum length of four to five atoms. The chain's electronic structure studied by STS revealed a strong dependence on the orientation of the chain with respect to the substrate orientation. But, also for this system, local investigations of the spin-related properties are, so far, lacking.

Finally, obtaining electrical signals of spin excitations of atoms on semiconductors is hampered by the gap in the density of states of the substrate that necessitates a relatively large stabilization bias for STS. Notwithstanding this difficulty, it has been possible to detect spin excitations of individual Fe atoms adsorbed to the (110) surface of InSb using IETS~\cite{Alex_Nature}. In this work, the dilute surface electron doping via the Fe atoms acting as donors induces a 2-D electron system at the surface which circumvents the problem opposed by the band gap. Most remarkably, the 2-D electron system interacts with the spins of the Fe atoms in an interesting fashion. Namely, it has been shown that, in an applied magnetic field, the $S=1$ Fe spins acts as spin filters for the substrate electrons that tunnel between the spin-polarized Landau levels of the 2-D electron system and the metal STM tip. This indicates a considerable exchange interaction between the electrons of the 2-D system and the Fe spins. Unfortunately, so far, also for this promising sample system no chains were prepared and studied using IETS or SP-STS.

%% file: SpinChainsCoupled/metals_superconductors.tex
\section{Spin Chains Strongly Coupled to the Substrate's Electron Bath}
\label{Coupled}

At low temperature, electronic excitations are the primary source of
spin de-excitation and decoherence. We saw in the previous chapter how
quantum properties of spin chains can be singled out and preserved by
decoupling the atomic spins from the electronic degrees of freedom
of the substrate. On the other hand, spin chains \nico{that are strongly
coupled to substrates with ubiquitous electronic states are also of
fundamental interest. In this case, the delocalized electronic states
can efficiently mediate spin-spin long-range interactions (RKKY) 
including a considerable Dzyaloshinskii-Moriya contribution. This
permits us to tailor the spin states by choosing the substrate material
and the interatomic distances in the chain.} Using heavy-element and/or
superconducting metals as substrates, intriguing emergent properties
like spin spirals or 1-D topological superconductors can be realized,
as we will review in the following.

\subsection{Spin Chains on Metallic Substrates}
\label{MetalSub}

\begin{figure}
\includegraphics[width=1.0\columnwidth]{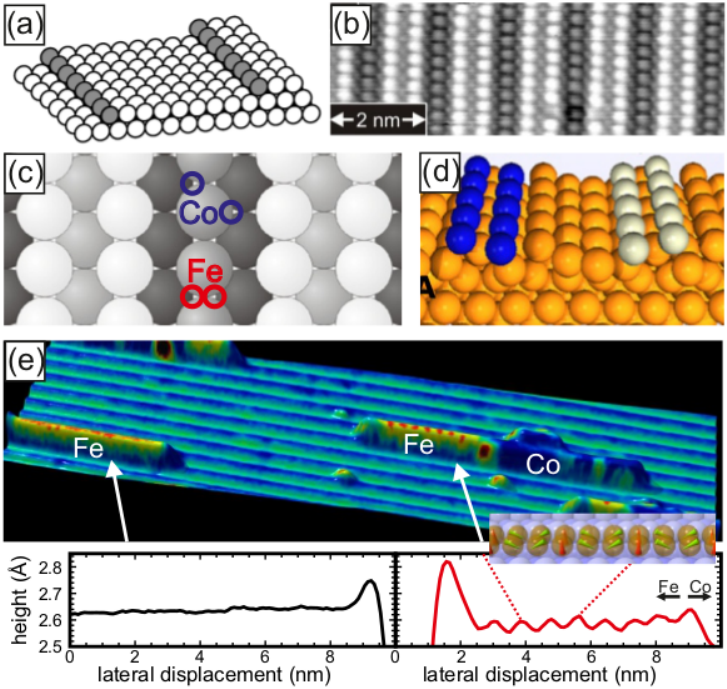}
\caption{\label{fig:metalsubstrate_Fig1}
Self-organized magnetic chains on metallic substrates. (a)~Sketch of monatomic Co chains attached to the step edges of a Pt(997) single crystal as reported in~\cite{Gambardella_2002}. (b)~STM topography with atomic resolution of the $(5\times1)$-Ir(001) surface and (c)~top view ball model of this reconstructed surface and characteristic adsorption sites (both adopted from~\cite{MenzelPRL2012}). (d)~Symmetry equivalent biatomic zigzag chains on the $(5\times1)$-Ir(001) surface as realized by Co (taken from~\cite{DupeNJP2015}). (e)~SP-STM topography colorized with the simultaneously obtained differential conductance signal of biatomic Fe and Co chains grown on $(5\times1)$-Ir(001); whereas the Fe chain on the left switches its magnetization frequently the magnetic spin spiral ground state of the right Fe chain (see sketch) is fixed by direct exchange coupling to the adjacent ferromagnetic Co chain (adapted from~\cite{MenzelPRL2012}).}
\end{figure}

\textit{Chains by Self-Organization.---}
An early prominent example of the preparation and investigation
of magnetic chains on metallic surfaces was Co chains grown by
self-organization on a Pt surface~\cite{Gambardella_2002}. The substrate
was a vicinal Pt(997) surface obtained by cutting a Pt crystal with a
misalignment of $6.45^\circ$ relative to the (111) plane. In this case,
rather narrow terraces with (111) orientation are formed and the deposited
Co atoms tend to bind at the step edges, forming long 1-dimensional Co
chains (Fig.~\ref{fig:metalsubstrate_Fig1}(a)). Using X-ray magnetic
circular dichroism (XMCD) the experiments showed that, at low enough
temperatures, these chains form a long-range-ordered ferromagnetic state
due to a rather large magnetic anisotropy. Interestingly the easy axis
is canted by 43$^\circ$ with respect to the (111) surface normal towards
the steps. 
\nico{Further physical insight
can be obtained by studying  the properties of single chains using
SP-STM.}

In addition to substrates with large miscut angles
as for Pt(997), also uniaxial surfaces can be used as templates for
the self-organized growth of chains. The growth of chains consisting
of magnetic atoms has been demonstrated for the reconstructed
$(5\times1)$-Ir(001) surface~\cite{Heinz2009} (see atomically resolved
STM image in Fig.~\ref{fig:metalsubstrate_Fig1}(b)), and different
kinds of biatomic chains have been studied regarding their magnetic
properties~\cite{MenzelPRL2012,DupeNJP2015}. The magnetic atoms can
adsorb on different adsorption sites leading to biatomic chains with
different symmetry (Fig.~\ref{fig:metalsubstrate_Fig1}(c)): the atoms can
sit in adjacent hollow sites preserving the symmetry of the underlying
substrate, or zigzag chains can be formed that break the mirror plane
within the chain axis (see sketches of the two resulting zigzag chains
in Fig.~\ref{fig:metalsubstrate_Fig1}(d)). 
Such zigzag chains are realized when Co is deposited and they were studied using SP-STM. 
Similar to the case of the Co chain attached to
the step edge of Pt(997), the Co biatomic chains on $(5\times1)$-Ir(001)
are ferromagnetic with an easy magnetization axis canted with respect
to the surface normal~\cite{DupeNJP2015}.

Biatomic Fe chains on the $(5\times1)$-Ir(001) conserve the symmetry of
the substrate, i.e.\ two orthogonal mirror planes, and the easy axis is
restricted to the high-symmetry directions. In contrast to the Co chains,
where at 8~K the magnetic anisotropy energy is sufficiently large to
suppress a thermally induced magnetization switching, the magnetic state
of the Fe chain switches on a time-scale that is much shorter than the
typical SP-STM signal detection time resolution of ms/pixel. Consequently,
at zero magnetic field, the spin-polarized contribution to the tunnel
current is averaged and vanishes, leading to a homogeneous signal along
the chain (see left chain in Fig.~\ref{fig:metalsubstrate_Fig1}(e)),
posing a challenge to the experimental investigation of such
one-dimensional magnetic chains.

In SP-STM measurements on biatomic Fe chains on $(5\times1)$-Ir(001)
at 8~K a magnetic signal is only detected when the thermal
fluctuations of the magnetic state are suppressed by either
direct exchange coupling to a stable magnet, as demonstrated in
Fig.~\ref{fig:metalsubstrate_Fig1}(e) with a ferromagnetic Co chain,
or by application of an external magnetic field~\cite{MenzelPRL2012}
(see below in Fig.~\ref{fig:metalsubstrate_Fig3}(c)). \nico{The observed
magnetic superstructure of about three atomic distances  originates from
a spin-spiral ground state. As revealed by DFT,} this magnetic
state results from a competition of direct (nearest-neighbor), indirect
(more distant neighbors via substrate), and antisymmetric exchange
interactions (Dzyaloshinskii-Moriya). The direct exchange interaction
is strongly ferromagnetic for the pairs perpendicular to the chains axis,
but only very small along the chains with a similar order of magnitude
as exchange between more distant spins and the Dzyaloshinskii-Moriya
interaction~\cite{MenzelPRL2012}. 

\begin{figure*}
\includegraphics[width=1.8\columnwidth]{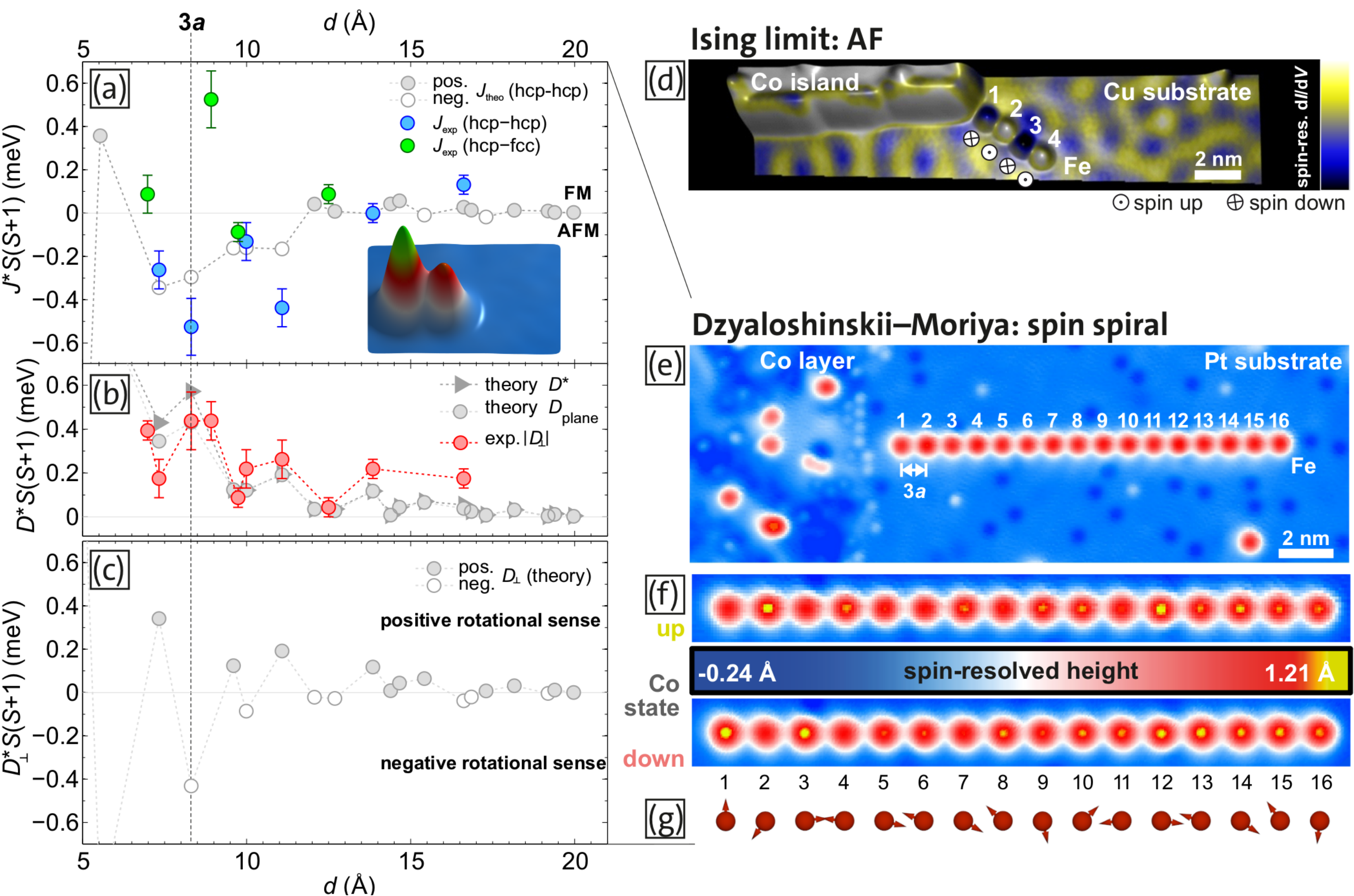}
\caption{\label{fig:metalsubstrate_Fig2}
STM-tip crafted magnetic chains on metallic substrates. (a,b)~Experimentally measured and DFT calculated Heisenberg (a) and Dzyaloshinskii-Moriya (b) components of the RKKY interaction between an Fe atom and an Fe-hydrogen complex on Pt(111) (see inset) as a function of their separation $d$. (c)~DFT calculated in-plane component $D_\perp$ of the Dzyaloshinskii-Moriya vector which determines the indicated rotational sense of the magnetization in the pair ((a-c) are adapted from~ \cite{Khajetoorians2016}). (d)~Spin-resolved image of a chain of 4 antiferromagnetically RKKY-coupled Fe atoms on Cu(111) which are stabilized by RKKY interaction to a magnetic Co island (adapted from~\cite{Alex_science}). The color reflects the spin-orientation of each atom in the chain, as indicated by the symbols. (e,f)~Spin-resolved images of a chain of 16 antiferromagnetically RKKY-coupled Fe atoms on Pt(111) which are stabilized by RKKY interaction to a magnetic Co layer (adapted from \cite{Steinbrecher2017}). In (f) the magnetization of the Co layer is magnetized up (top) and down (bottom) resulting in the alignment of the spin of the first chain atom (up and down). (g)~Sketch of the approximate spin orientations of the Fe chain atoms.}
\end{figure*}

\textit{Chains by Atom Manipulation.---}
The self-\nico{organization} technique has the advantage of using
thermodynamics to achieve a reproducible way of creating a large number
of quasi infinitely long chains. However, it is also totally dependent
on the patterning of the substrate. Therefore, the distance between the
atoms in the chain is rather fixed by the substrate properties. The mutual
exchange interactions between the atoms in the chain cannot be varied
easily from the dense, direct-exchange dominated regime into the dilute,
indirect itinerant-electron exchange mediated, or so-called RKKY dominated
regime. Finally, it cannot be used to create individual spin chains of
an absolutely well defined number of atoms. A different approach is the
creation of spin chains by STM-tip induced atom manipulation which is
illustrated in Fig.~\ref{fig:metalsubstrate_Fig2}.

Using SP-STS~\cite{Zhou2010,Alex_NatPhys} or
IETS~\cite{Khajetoorians2016}, the magnetizations or excitations,
respectively, of two atoms of the desired chain material in an
RKKY-coupled pair can be measured as a function of an external
magnetic field. By fitting the data to models using effective spin
Hamiltonians \nico{(Sec.~\ref{Spin_Ham})} it is possible to extract
the isotropic ($J_{ij}$) and Dzyaloshinskii-Moriya ($\vec{D}_{ij}$)
components of this pair-wise RKKY interaction. An example is shown in
Fig.~\ref{fig:metalsubstrate_Fig2}(a-c) for an Fe atom and an Fe-hydrogen
complex on Pt(111) with increasing separation between them in comparison
to DFT. These maps of distance dependent exchange interactions can then
be used in order to tailor and build artificial dilute chains of magnetic
atoms on different substrates (Fig.~\ref{fig:metalsubstrate_Fig2}(d-g)).

For the example of Fe atoms on Cu(111)
(Fig.~\ref{fig:metalsubstrate_Fig2}(d)), it was found that
the longitudinal magnetic anisotropy $D$ of the atoms is the
dominant energy scale and about an order of magnitude larger than
$J_{ij}$~\cite{Alex_NatPhys}, while $\vec{D}_{ij}$ is negligible
because of the relatively light substrate and consequently weak
spin-orbit interaction. Therefore, this system behaves like an
Ising system. By choosing an appropriate interatomic distance, it
was possible to stabilize the N\'{e}el state in artificial chains
with different number of atoms. A detection by SP-STM was possible
by either coupling the first atom to a ferromagnetic island via RKKY
interactions~\cite{Alex_science} (Fig.~\ref{fig:metalsubstrate_Fig2}(d))
or, for odd-numbered chains, by stabilizing one state in a
weak external magnetic field~\cite{Alex_NatPhys} (see below,
Fig.~\ref{fig:metalsubstrate_Fig3}(a)).

For the heavier substrate Pt(111) with stronger spin-orbit
coupling (Fig.~\ref{fig:metalsubstrate_Fig2}(a-c)), the two
RKKY-contributions $J_{ij}$ and $\vec{D}_{ij}$ are of similar strength
and comparable to the strength of the longitudinal magnetic anisotropy
$D$~\cite{Khajetoorians2015}. Due to symmetry reasons, the main
component $D_\perp$ of $\vec{D}_{ij}$ lies in the surface plane and is
perpendicular to the displacement vector between the two atoms. Therefore,
for weak magnetic field, the pair is in a non-collinear spin state,
and by adjusting the distance $d$, the sign and strength of $D_\perp$,
and thereby the sense and angle of rotation of the spin from one to
the other atom can be tailored. By RKKY-coupling the first atoms of an
artificial dilute 16-Fe atom chain of appropriate atomic distance to
a ferromagnetic Co layer (Fig.~\ref{fig:metalsubstrate_Fig2}(e)),
it was indeed possible to stabilize a spin-spiral state
(Fig.~\ref{fig:metalsubstrate_Fig2}(g)) as proven by the change in the
spin-resolved apparent height of the chain atoms when reversing the Co
layer (Fig.~\ref{fig:metalsubstrate_Fig2}(f)). These results demonstrate
that the knowledge of the distance-dependent RKKY-interaction in
combination with STM-tip induced manipulation allows for a high level
of versatility and tunability of spin chains of various elements on
various substrates.

\begin{figure}
\includegraphics[width=1.0\columnwidth]{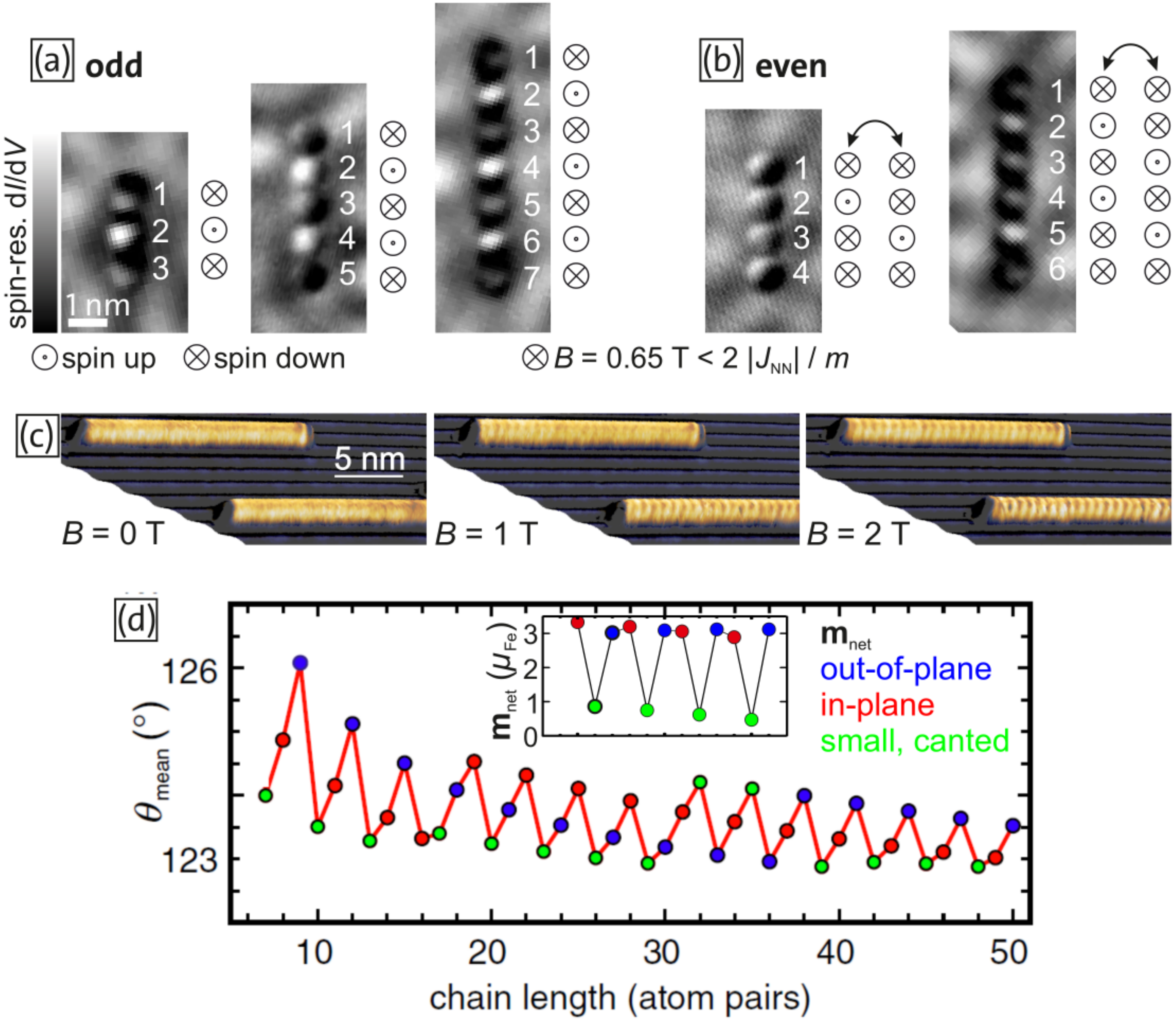}
\caption{\label{fig:metalsubstrate_Fig3}
Parity effects in STM-tip assembled and self-organized spin chains on metallic substrates. (a,b)~Spin-resolved differential conductance images of several odd- (a) and even- (b) numbered artificially constructed Fe chains on Cu(111) in the indicated magnetic field (adapted from \cite{Alex_NatPhys}). The resulting N\'{e}el states (odd case) and the two states between which the chain fluctuates (even case) are given on the right side of each image. (c) Self-assembled biatomic Fe chains on $(5\times1)$-Ir(001) as a function of external out-of-plane magnetic field. (d) Results from micromagnetic simulations showing the chain-length dependent variation of the mean angle and the net magnetic moment ((c) and (d) taken from \cite{MenzelPRL2014}).}
\end{figure}


\textit{Parity Effects.---}
Using self-assembled as well as manipulated antiferromagnetic
or non-collinear chains, interesting effects on the number of the
atoms in the chain, so called parity effects, have been studied. Such
parity effects are generally based on the dependence of the strength
or orientation of the net magnetic moment of the chain on the \nico{parity of the} number
of the chain atoms. \nico{For example,} an antiferromagnetic Ising chain of an
odd number of atoms has a nonzero net magnetic moment. 
In a weak homogeneous magnetic field which is small enough such
that the Zeeman energy cannot break the nearest neighbour exchange
interactions ($B < 2\left|J_\textrm{NN}\right|/m$ with the nearest
neighbor interaction $J_\textrm{NN}$ and the modulus of the magnetic
moment of each atom, $m$), the stabilization of the N\'{e}el state
is expected. This state was indeed observed in odd-numbered Fe
chains on Cu(111) using SP-STS in a weak external magnetic field
(Fig.~\ref{fig:metalsubstrate_Fig3}(a))~\cite{Alex_NatPhys}. In
contrast, for the even-numbered Ising chains, the net magnetic
moment is zero, such that a homogeneous magnetic field cannot
stabilize the N\'{e}el state. Depending on whether the system is in
the classical or quantum mechanical limit, the resulting state either
fluctuates between two degenerate classical states, or, respectively,
is a quantum superposition. In the former case, the fluctuation
is typically much faster than the time resolution of conventional
SP-STS, resulting in a loss of the magnetic contrast on chains with
even number of atoms, as indeed observed for Fe chains on Cu(111)
(Fig.~\ref{fig:metalsubstrate_Fig3}(a)). In this case, the N\'{e}el
state can still be stabilized by a local RKKY-exchange field acting on
the end of the chain, e.g. by RKKY-coupling the chain end to a stable
ferromagnetic island (Fig.~\ref{fig:metalsubstrate_Fig2}(d)). For
the quantum mechanical limit, i.e.\ spin chains largely decoupled
from the substrates electron bath (see section \ref{Decoupled})
even-odd effects have also been studied. For Mn chains on Cu$_2$N (see
Fig.~\ref{fig:Mn_chain}) they manifest in the presence or absence of
the singlet-triplet excitation for the even- and odd-numbered chains,
respectively. For Co chains on Cu$_2$N they are visible as a qualitative
difference between the magnetic field dependent IETS data of even-
and odd-numbered chains as shown in Fig.~\ref{fig:xxz}(c).

An even-odd effect has also been proposed for short Mn
chains that have been manipulated on a ferromagnetic Ni(110)
substrate~\cite{holzberger_2013}. Again, there is antiferromagnetic
coupling between the chain atoms. However, in this case, the chain atoms
are additionally ferromagnetically exchange coupled to the substrate which
is magnetized along the chain axis, resulting in a homogeneous exchange
field along the chain. Here, DFT predicts a N\'{e}el state (collinear)
oriented along the chain axis for the odd-numbered chains. For the
even-numbered chains, the calculations predict a fluctuation between
two degenerate non-collinear ground states which are generated by the
exchange field from the substrate and the magnetic anisotropy. When
the magnetic atoms of a chain on a magnetic substrate are positioned at
larger distances the exchange coupling along the chain can be reduced
and the magnetic moments of the chain mimic the magnetic structure of
the underlying substrate. In the case of Co atoms on a Mn monolayer on
W(110) this structure is a spin spiral~\cite{Ser2010}.

When the magnetic state of the chain itself is non-collinear,
the even-odd effects generally become more complex. In the case of
biatomic Fe chains on $(5\times1)$-Ir(001) the differences between the
magnetism of chains with varying lengths manifest in the magnetic field
dependence of the amplitude of the spin contrast (see the chains in
Fig.~\ref{fig:metalsubstrate_Fig3}(c))~\cite{MenzelPRL2014}. Because the
period of the spin spiral is nearly 3 atoms, micromagnetic simulations
using the DFT parameters have identified three different types of
chains that alternate: they can be classified according to the size
and direction of the net magnetic moment of the chain. Depending
on the number of atomic pairs of the chain, the net magnetic moment
is either large and perfectly out-of-plane or perfectly in-plane,
or it is small and has no characteristic direction (see graph in
Fig.~\ref{fig:metalsubstrate_Fig3}(d)). Even though the magnetic period is
given by the competition of the magnetic interactions this parity
effect becomes also evident in the mean angle between the Fe atom pairs
(Fig.~\ref{fig:metalsubstrate_Fig3}(d)). However, the distortions are
very small; note that due to the small deviation from $120^{\circ}$
between adjacent magnetic moments there is a long-range pitch in the
succession of the three chain types.

\subsection{Spin Chains on Superconducting Substrates}
\label{Shiba}

\textit{1-D Topological Superconductivity.---}
Recently, chains of magnetic atoms on superconducting substrates are
investigated regarding the possibility to achieve \nico{the} so-called topological
superconductivity in one dimension \nico{that goes} along with the
emergence of Majorana bound states (MBS) at the chain's ends. The
realization of MBS in solid state systems is strongly desired because of
their peculiar statistical properties. If two MBS are exchanged, they
produce a non-trivial phase that is related with the topology of the
crossing trajectories. Kitaev showed that operations with MBS could be
used to develop new schemes of quantum computation, reducing operational
errors~\cite{Kitaev}. Additionally, fermion-parity conservation of
the topological superconductor removes problems of decoherence that
usually limit quantum computations~\cite{Kitaev,Rainis_2012}. Moreover,
the formation of the MBS is a strongly non-local effect, as they always
come in pairs, one on each end of the chain. Therefore, MBS are expected
to be largely immune against local perturbations. These findings have
inspired many theoretical works (for a small sample of review topics in
this subject please consult~\cite{Nayak_2008,Stern_2010,Alicea_2012,Beenakker_2013,Stern_2013,Tokura_2017}).

In an instructive toy model, Kitaev showed that MBS will appear
at the ends of a one-dimensional and effectively spin-less p-wave
superconductor~\cite{Kitaev_2001} that is adiabatically connected to a
1-D topological superconductor~\cite{Pientka_2015}. Usually, in elemental
superconductors, the orbital part of the wave function of the Cooper pairs
is described by an s-wave \nico{that is even under particle interchange,
and the spin part by the odd singlet state that warrants an odd wave
function}. In p-wave superconductors, the orbital part of the wave
function is a p-wave \nico{that is odd under particle interchange, and
the spin part is then even (the Cooper-pair spin is $S=1$)}. The latter
enables pairing of effectively spin-less electrons, i.e.\ electrons of
the same spin-type, such that all low-energy excitations are spin-less
and the electronic spin can be effectively left out.  Unfortunately,
the realization of p-wave superconductivity is still controversially
discussed \nico{for the only proposed bulk candidate (Sr$_2$RuO$_4$).}

\nico{There are two present experimental approaches to}
 realize a 1-D topological superconductor.  In the first
approach, semiconductor nanowires with Rashba-type spin-orbit interaction
are coupled to an s-wave superconductor and a magnetic field
is applied in a direction perpendicular to the spin-orbit effective
field~\cite{Mourik_2012, Albrecht_2016, Lutchyn_2018}.  The rational
behind these experiments is to
 combine the three following ingredients~\cite{Pientka_2015}:
(i) Proximity coupling of a 1-D electron system to a bulk s-wave
superconductor in order to \nico{transfer superconductivity to the
nanowire, circumventing the Mermin-Wagner theorem that implies the
impossibility of superconductivity in one dimension~\cite{Mermin-Wagner}. (ii) A Zeeman
field needed to spin-polarize the electron system such that
it is essentially spin-less. (iii) Strong spin-orbit coupling and/or
the impressing of helical spin states in order to enable  Cooper
pairing of  electrons.} 
Different signatures of MBS have been observed in such
structures. In April 2018, the Delft group announced that the zero bias
conductance in their wires reached the predicted limit of the quantum
of conductance~\cite{Zhang_2018}, strong evidence
for the existence of MBS at the ends of their nanowires.

A second approach is to couple chains of magnetic atoms to s-wave
superconductors. Here, the above ingredient (ii) can
be circumvented by using ferromagnetic spin chains. Also, ingredient (iii) can be achieved
either by using materials (chain or superconductor) that have
an intrinsically large spin-orbit interaction~\cite{Li_2014},
or by inducing non-collinear, e.g. spin-spiral states, in the
chain~\cite{Martin_2012, Braunecker_2013, Beenakker, Pientka_PRB_2013,
Klinovaja_2013, Nad2013, Vazifeh_2013, Schecter_2016} \nico{that also}
overrides ingredient (ii). The constituents of such spin chains,
i.e.\ the individual magnetic atoms coupled to the surface of the
superconductor, already induce states in the energy gap. These
states are usually named Yu-Shiba-Rusinov (YSR) states after their
discoverers~\cite{Yu_1965,Shiba_1968,Rusinov_1969}. In view of their
relation and importance for MBS, the properties of the YSR states have
been studied in detail\nico{. Let us review them briefly} before
we turn to the description of the few experimental realizations of
spin-chain on superconductor systems in the so-called Shiba chain and
wire limits~\cite{Pientka_2015} where the latter have revealed strong
indications for MBS.

\textit{YSR states.---}
YSR states originate from the weakening of the binding of Cooper
pairs induced by the magnetic atom on an s-wave superconductor.
In order to inject or extract a single
electron from the superconductor, the electron energy has to be larger
than $\Delta$, which can be interpreted as the binding energy of the
Cooper pair~\cite{deGennes,Tinkham}. A magnetic atom in a superconductor
can scatter electrons if their energy is larger than $\Delta$, but it can
also scatter Cooper pairs. If we assume that $\Delta$ is large enough,
spin-flips by the magnetic impurity can be safely ignored, and, in first
approximation, the effect of the impurity on the conduction electrons can be regarded as that of an exchange field. The exchange
field will act on the two electrons of the s-wave Cooper pair in a different
way, depending on their spin, hence weakening the binding energy of the
Cooper pair. As a consequence, it can be easier to break this Cooper
pair and the energy to inject or extract an electron will be less than
$\Delta$. This can create a state in the energy gap of the superconductor,
the YSR state.

\begin{figure}[t!]
\includegraphics[width=1.0\columnwidth]{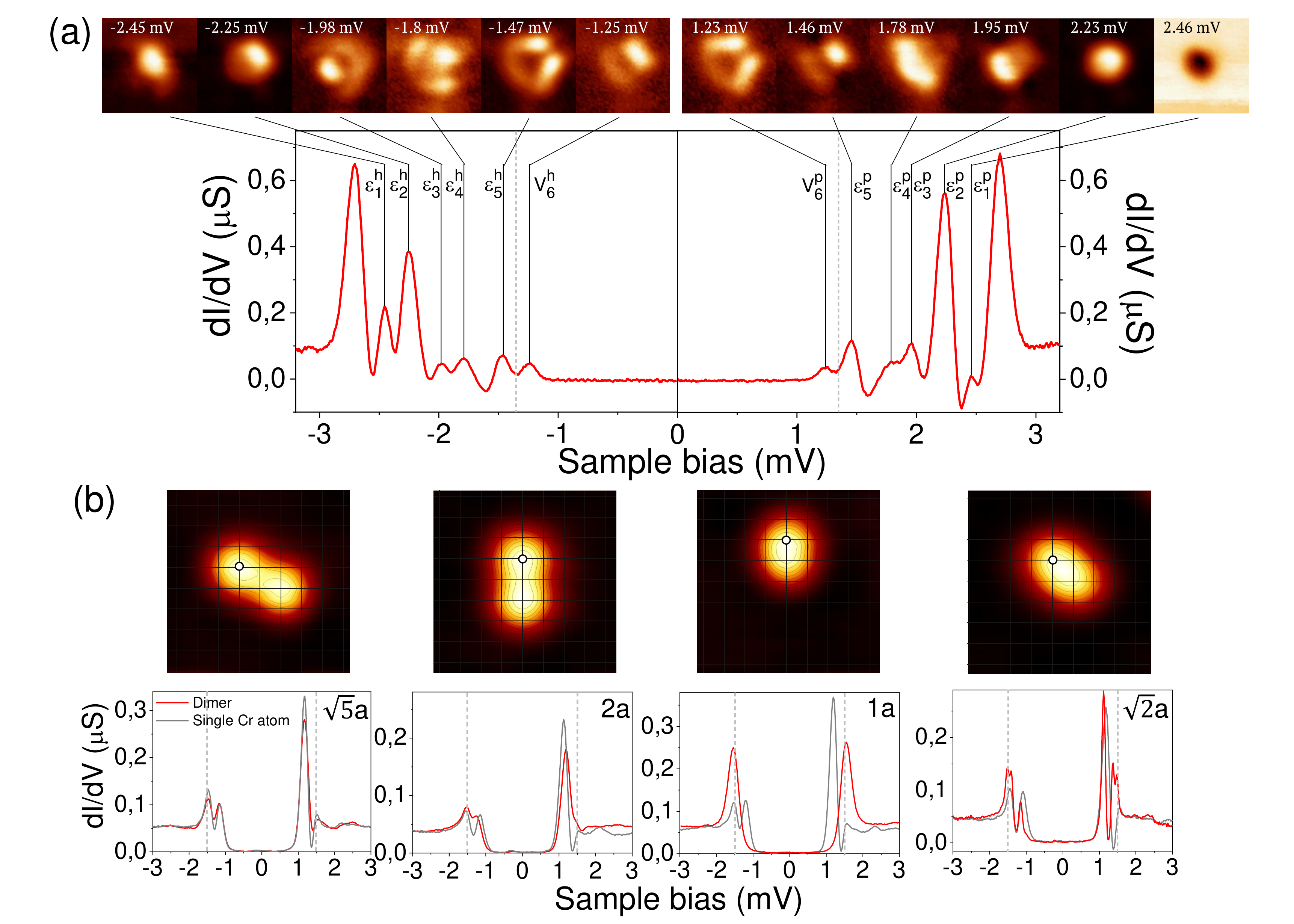}
\caption{\label{Shiba_dimer}
YSR states in single Cr atoms on Pb(111) and Cr dimers on a $\beta$-Bi$_2$Pd superconductor.
(a) Multiple YSR states are shown over a single Cr atom on Pb(111), originating from a hybridized orbital structure of Cr with nearest neighbor Pb atoms.
Differential conductance maps at the peak positions of the YSR states show the hybridized orbital features (reproduced from~\cite{Choi_2017}).
(b) Dependence of YSR states on the spacing of Cr atoms in dimers on $\beta$-Bi$_2$Pd.
Separations from left to right: $\sqrt{5}$, 2, 1, and $\sqrt{2}$ by unit cell distances. All dimers were formed by STM-tip induced atomic manipulation. The top panels show the topographic images of the dimers, the bottom panels the corresponding differential conductances measured on one of the two atoms in the dimer (red lines) and on a single atom for comparison (grey lines) (reproduced from~\cite{Choi_2018}).}
\end{figure}

After the first experimental verification of YSR states of individual
atoms using STS~\cite{Yazdani_1997}, numerous experimental studies
have been performed, revealing effects due to the orbital structure of
the atoms~\cite{Ji_2008,Ruby_2016,Choi_2017}, due to the magnetic
anisotropy of the atom~\cite{Hatter_2015}, due to a reduced dimensionality
of some superconductors~\cite{Menard_2015}, due to the competition between
Kondo screening and Cooper pairing~\cite{Franke_2011,Bauer_2013}, and due
to the spin-polarization of the YSR state~\cite{Jens}. Because of the
orbital structure of the magnetic atom, there are spatial variations
of the exchange field produced by the atom, which induce a marked
shape of the spatial distribution of the YSR state. This has been
recently revealed in STM measurements of Cr and Mn impurities on Pb
surfaces~\cite{Ruby_2016,Choi_2017}. Fig. \ref{Shiba_dimer} (a)
shows the multiple YSR states originating from the hybridization of the
orbitals of the Cr atom with those of the surrounding Pb atoms.

Using SP-STS, the theoretically predicted spin-polarization of the
YSR states of individual magnetic atoms on a superconductor has been
experimentally detected~\cite{Jens}. To this end single Fe atoms on the
(3$\times$3) oxygen reconstruction on Ta(100) have been magnetized in a
small external magnetic field which was weaker than the critical field of
the superconducting substrate. The SP-STS data, taken with a magnetic tip
that was thoroughly characterized regarding its spin-polarization on a
reference system, showed the expected sign change in the spin-polarization
between the electron and hole parts of the YSR state~\cite{Jens}.

\textit{Shiba chain limit.---}
When dilute arrays, e.g. chains, of transition metal magnetic atoms
are assembled on a superconductor, the YSR states start to overlap,
and can eventually form so-called Shiba bands. Here, it is assumed
that the atoms are sufficiently separated, such that direct hopping
between the atom's d-levels can be neglected, which is the so-called Shiba chain
limit. If the Shiba band is close to the Fermi level and broad enough
to overlap with it, topological superconductivity can
evolve~\cite{Pientka_2015}. Some recent works, therefore, probe and
characterize the interactions between YSR states with respect to
the formation of such a Shiba band \cite{Liljeroth_2017,Choi_2018,
Ruby_2018}. The investigation of pairs of Mn atoms on Pb(001) reveals the
formation of symmetric and antisymmetric combinations of YSR states which
is studied as a function of the orientation of the pairs with respect
to the orbital shape of the individual YSR states \cite{Ruby_2018}. In
the work of ~\cite{Liljeroth_2017}, the interaction of the YSR states
of magnetic molecules on a NbSe$_2$ substrate is investigated. Here,
the formation of coupled YSR states is visible for relatively distant
molecules. This is facilitated by \nico{the large spatial extent of the YSR
states due to the 2D character of} superconductivity in this
material~\cite{Frindt_1972,Gerbold_2015,xi_2015,Ugeda_2016}.

Choi \textit{et al.}~\cite{Choi_2018} studied pairs of Cr atoms
on $\beta$-Bi$_2$Pd, a type-II superconductor with critical
temperature of 5.4 K~\cite{Imai_2012} that has been shown to contain
topological surface states~\cite{Sakano_2015} characterized by a unique
superconducting gap despite its multiband structure~\cite{Herrera_2015}
(see Fig.\ref{Shiba_dimer} (b)). A single Cr atom on $\beta$-Bi$_2$Pd
produces YSR states~\cite{Choi_2018}.  When a second Cr atom was
approached along the [100] direction of the surface, the YSR states
shifted and broadened. As the distance was shortened, there was a clear
shift to higher energy, driving the YSR state towards the quasi-particle
continuum edge. At very short distances, leading to the formation of a
Cr$_2$ dimer, the YSR peak disappeared into the continuum. These data
were interpreted as the disappearance of the localized magnetic moment
over the Cr--Cr system, which implies an antiferromagnetic interaction
between the two magnetic moments.  For pairs oriented along the [1-10]
direction, the YSR states splitted, indicating a hybridization of the
YSR states consistent with a ferromagnetic coupling of the Cr magnetic
moments~\cite{Choi_2018,Flatte_2000}.

\begin{figure}
\includegraphics[width=1.0\columnwidth]{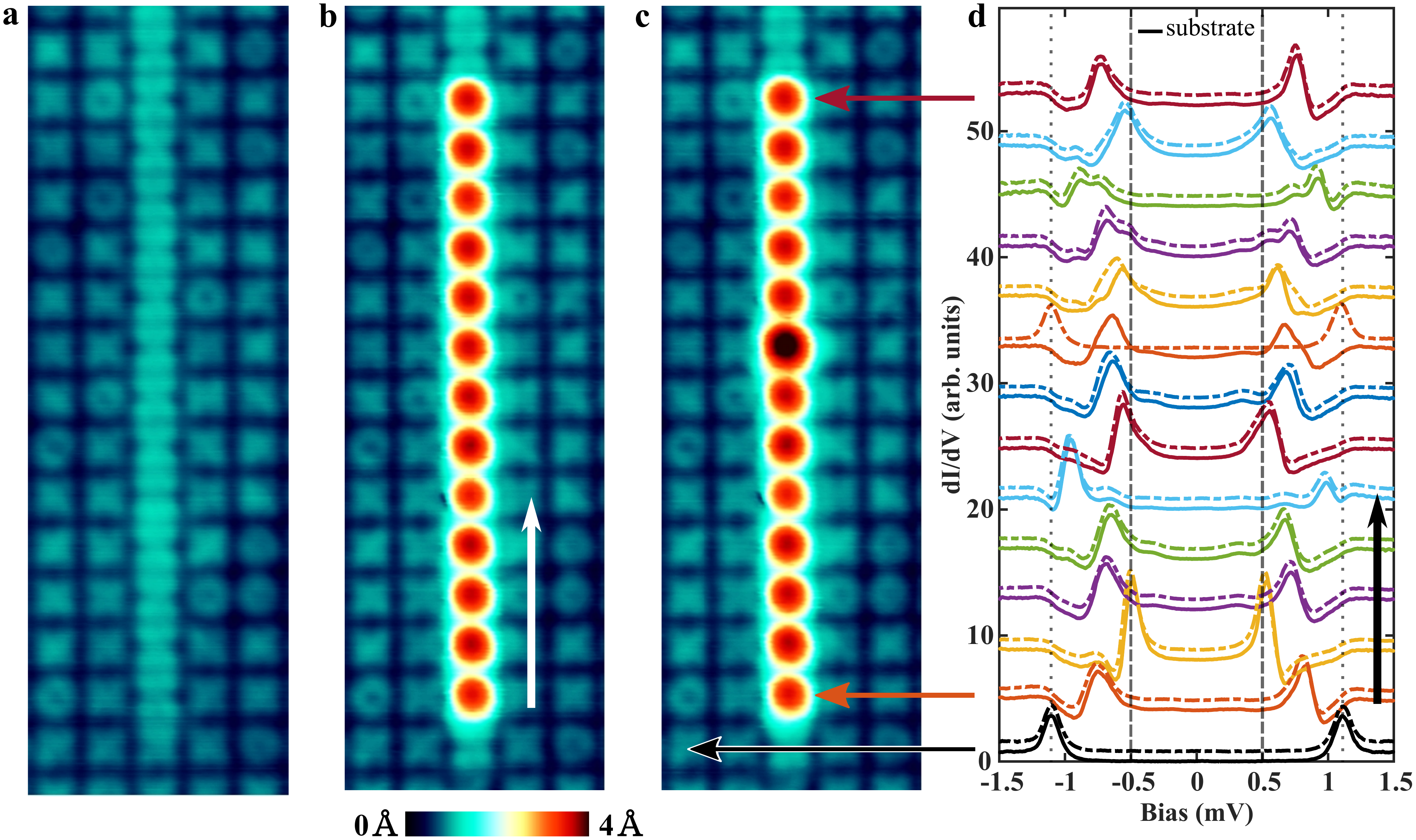}
\caption{\label{Fe_chains_Ta}
STM-tip assembled chain of weakly coupled Fe atoms on Ta(100)-(3$\times$3)O. (a) Assembled chain of subsurface interstitial Fe atoms. (b) 13 Fe atoms have been assembled on the surface along the subsurface atoms, such that each Fe atom is linked by a subsurface Fe atom. (c) The sixth atom from the top has been switched into a non-magnetic state. (d) Comparison of differential conductance spectra taken with a superconducting tip on every Fe adatom and on the substrate, with all atoms in the magnetic state (solid lines) and the sixth atom in the nonmagnetic state (dashed lines). The dashed (dotted) vertical lines indicate the sample Fermi level (sample coherence peak) (adapted from~\cite{Kamlapure_2018}).
}
\end{figure}

Along these lines, artificial chains of magnetic
atoms with YSR states were assembled on a superconducting substrate
using STM-tip induced atom manipulation. Kamlapure \textit{et
al.}~\cite{Kamlapure_2018} investigated the coupling of the YSR
states in artificial chains of Fe atoms assembled on the (3$\times$3)
oxygen reconstruction of Ta(100) (see Fig. \ref{Fe_chains_Ta}). While
pairs of Fe adatoms showed a negligible interaction of the YSR states due
to a relatively large  Fe--Fe distance, the interaction was increased by
the manipulation of subsurface interstitial Fe atoms in the center between
the two Fe adatoms, as proven by the shift of the YSR states. Motivated by
this effect, chains of Fe adatoms and subsurface interstitial Fe atoms
were assembled (Fig. \ref{Fe_chains_Ta}a,b) and investigated concerning
the formation of a YSR band (Fig. \ref{Fe_chains_Ta}c,d). Even though
there is a considerable interaction between the YSR states as visible
from the change in the YSR state energy when one of the chain atoms is
switched into a nonmagnetic state (b to d), the YSR state energies are
distributed inhomogeneously along the chain (Fig. \ref{Fe_chains_Ta}d)
indicating considerable electronic disorder, which prevents the formation
of a YSR band. So far, an experimental realization of a clean YSR band
in a chain of dilute magnetic atoms on a superconductor, i.e. within
the Shiba band limit, is still missing.

\textit{Wire Limit.---}
When the transition metal atoms in a chain are more densely packed, direct
hopping between the d-levels of the atoms can no longer be neglected. If the
resulting d-band crosses the Fermi level, it has to be considered in the
description of the low-energy phenomena. This limit is
called the wire limit~\cite{Pientka_2015}. Two systems reported in the
literature so far show evidence for MBS at the ends of spin chains in
this limit. Nadj-Perge \textit{et
al.} investigated Fe chains attached to clusters on the Pb (110) surface.
These spin chains are formed after
room-temperature deposition of Fe, followed by annealing~\cite{Ali2014}. With the
help of DFT, the most probable structure of the chain was found to be a
three-layer Fe zigzag chain partially submerged in the Pb. Spin-resolved
STS revealed a contrast consistent with a ferromagnetic alignment of
the topmost Fe atoms in the chain, and it was concluded with the help
of DFT, that the chain is in a ferromagnetic state. STS revealed a zero
bias peak within a length of 1 to 2 nm from the end of the chains
which was interpreted as the signature of a MBS. The zero bias peak
in this system was reproduced by two other groups~\cite{Ruby_2015b,
Pawlak_2016}. Ruby \textit{et al.} pointed out that it is only present
for some of the chains, and moreover, as revealed by higher resolution
studies, that the zero bias peak overlaps with a low-energy resonance
that was tentatively attributed to the coherence peak of the induced
topological gap~\cite{Ruby_2015b}. A study performed at even lower
temperatures~\cite{Feldman2016} on the same system showed that the
zero bias peak has no detectable splitting, is particle-hole symmetric
for some of the chains and asymmetric for others, and has a significant
spectral weight in the substrate. Finally, spin-resolved STS of the zero
bias peak~\cite{Ali2017} reveals signals that are symmetric with respect
to bias reversal, in contrast to the antisymmetric signals observed for
YSR bands. The absence of splitting and particle-hole symmetry strongly
support the MBS interpretation of the zero bias peak.

Experiments of
Co chains on Pb (110) with a similar morphology and signature of
ferromagnetic order as for the case of Fe chains, yielded
 a zero-bias signal delocalized along the chain. The lack of localization
at the edges precludes the presence
of MBS~\cite{Ruby_2017}. Tight-binding
calculations for this system indicated an even number of Fermi level
crossings of the spin-orbit-split bands, in contrast to an odd number
found for the Fe chain system~\cite{Ruby_2017}, which would explain the
absence and presence of a topological phase for the Co and Fe systems,
respectively.

The experimental difficulty in finding a consistent signature of
MBS for some of the chains of the Fe/Pb(110) system which have been
investigated~\cite{Ruby_2015b, Feldman2016}, most probably originates in
an imperfect atomic structure, as evident from the variance in topographic
features found at the ends of the chains. This problem can be
circumvented by the investigation of artificial chains that are built by
STM-tip induced manipulation. Unfortunately, for the system Pb(110), this
procedure turned out impossible. However, Kim \textit{et al.} were able to
manipulate virtually atomically perfect chains of several tens of Fe atoms
on the (0001) surface of the strong-spin-orbit coupling superconductor
Re~\cite{Kim_2018}. Spin-resolved STS of chains of different numbers of
Fe atoms positioned on the close-packed hcp hollow sites of the Re(0001)
surface reveals a \nico{non-collinear magnetic} state. STS furthermore shows zero
bias conductivity localized in a region of half a nm width at both
chain ends, which disappears for chains with less than 9 Fe atoms. 
\nico{With the help of tight-binding model calculations based on parameters obtained
from ab-initio calculations, which predict that the chains are in the topologically superconducting state, 
the zero bias conductivity was interpreted as an indication for MBS at the chain ends.}
Manipulation of a
single-atom defect to the end of a short Fe chain interestingly revealed
that zero-bias peaks can be generated by such defects, stressing that
a full control of the chain composition is essential in order to rule
out trivial effects inducing zero bias peaks, that can potentially be
\nico{mistaken for} MBS.

%% file: OUTLOOK/outlook.tex
\section{Outlook}

Spin chains are the paradigm of quantum phenomena. Entanglement,
correlation, decoherence are properties inherent to spin chains. In this
Colloquium, we have shown an  overview of these phenomena in different
contexts. We specifically focused on scanning probes that,
besides atomic manipulation, {permit us to obtain detailed spectroscopic information on individual chains of precise length and composition.}

Complementary to other techniques such as atom traps or molecular
crystals, the STM manipulation of atoms on surfaces offers an extremely controlled
way of creating structures with tailored properties. The substrate
is the big constraint in these systems, which on the one hand, gives to environmental
perturbations on the properties of the spin chain. On the other hand, 
the use of substrates makes it possible to eventually encapsulate
and create devices, giving us ideas on how to create a useful technology
out of the quantum properties of entangled spins.

The approaches presented in this Colloquium {are related to many exciting research areas such as quantum information science.}
Let us briefly mention some of the interesting
connections.

\textit{Spintronics}.- The studies we have presented here bear direct
relation to the possibility of using spin instead of charge in solid-state
devices. Atom manipulation grants new capabilities to creating devices
with atomic precission. The rich spectra of spin chains and the different
ways to access them via electronic currents that we have presented
in this Colloquium show that indeed operation can be performed at
the atomic level in spins conveniently coupled to other spins and
decoupled from the degrees of freedom of the substrate. Spin chains
have been used to realize all-spin based logic operations~\cite{Alex_science}, to serve as tiny storage units of memory when
conveniently arranged using antiferromagnetic couplings~\cite{Loth_2012},
and inelastic effects have been shown to be an effective way of inducing
spin torque~\cite{Palacios,Loth_2010,Khajetoorians2013}.  We can easily envisage new
applications of spin chains in spintronic devices by using resonant
excitation of spin in time-dependent approaches~\cite{Baumann_2015}, or
combining them with the rich world of semiconductors (see section~\ref{SemiSub}) to
produce new devices.  Indeed, a new type of spin-based transistor has
been suggested \cite{Marchukov}. These authors show that by switching on
and off the entanglement with parts of the spin chain \nico{using
local spin excitations,} a spin-based transistor can be achieved.

\textit{Quantum communication}.- 
Using quantum mechanics to encode information and process it is a
tantalizing field with enormous possibilities.  Recent suggestions
show that information can be indeed transmitted with high fidelity
in spin chains.  In \cite{Khaneja} an Ising spin chain serves as
transmitting line of radio-frequency pulses that drive single-spin
information.  In \cite{Sougato} a Heisenberg spin chain is used by
putting to work its excitation spectrum.  There, it is shown that an
excitation in one extreme of the chain is partially transmitted to
the other edge of the chain. These results are regardless of the sign
of the Heisenberg coupling because they are a consequence of the full
spectrum of excitations. Perfect transmission has been proved to take
place in \textit{customized} spin chains. \cite{Landahl} prove that
perfect fidelity over long distance is obtained when using an XY-coupled
spin chain and also a Heisenberg spin chain where the couplings are
modulated by an external magnetic field.  Karbach and Stolze~\cite{Stolze} 
show, that, by tuning the parameters of the spin chain one can
actually obtain perfect transmission. Contrary to intuition, there is a
full class of inhomogeneously coupled chains that even allows for small
variations along the chain, permitting the transmission of information with
perfect fidelity. Karbach and Stolze~\cite{Stolze} further shows that transmission over
considerable distances can be achieved at arbitrary temperatures for
genuinely entangled states, giving rise to many technological options.
Quantum fluctuations can be  put to work in transmitting information
in an effective way \cite{Banchi}. Indeed, a new proposal uses a small
external magnetic \nico{field to} tune a given quantum phase of
the connecting spin chain and in this way control the transmission of
information \cite{Banchi}.

\textit{Quantum computing}.- Qubits and operations on qubits need to
be performed within the quantum coherence time. Using a solid device
is probably a difficult strategy due to the large number of degrees of
freedom and of interactions that will necessary perturb the acting
spins. Decoupling the spins is a strategy that seems to be working to
have access to these quantities with the STM, thus opening the door to
applications of solid-supported spin chains in quantum computing. We have
seen in the cases studied so far, that the solid and more generally,
the environment, becomes part of the quantum system. Typically this
is not beneficial because it can lead to faster decoherences and
other effects that destroy the superposition states needed for quantum
computation. However, the substrate can be beneficial. This is clearly
seen in the case of a superconducting substrate where it is the substrate
 that develops extraordinary topological properties leading
to Majorana edge states that can be potentially used for topological quantum
computing. Furthermore, superconductors have gaps that partially
decouple the spin chain, reducing excitation and other undesirable
phenomena. Choosing the substrate is an important aspect of the future
developments of spin chains. \cite{QI,QC}

\textit{Quantum simulations}.- This is a fascinating field where
atom traps are making big progress as we have briefly \nico{mentioned} in
the Introduction. Spins on surfaces can also be used by experimentally revealing the
solutions to model Hamiltonians that represent the behavior of matter
on the very-low-temperature scale. \nico{Indeed, the rich world of spin-based
Hamiltonians that have seen the light since the introduction of the
Bethe ansatz and integrable models based on spin chains gives us new methods to undertake the exploration of quantum matter. 
The study of
these systems with the new tools offered by the STM is an intriguing field with a huge perspective for future research.}
\cite{QS,QA}

\textit{Quantum sensors}.- The limits of metrology have been further
extended by the use of quantum measurements. Smaller quantities are
accessible using quantum effects thanks to the interference aspects of
superposition states. \nico{A recent experiment using electron-spin resonance
with an STM (ESR-STM), shows that atomic spins  can be used as extremely
precised and sensitive sensors~\cite{Natterer_2017}. The experiment
uses the shift in the electronic current resonance peak when the STM
tip is ontop of an Fe atom, as a detector of the magnetic dipolar interaction
with a nearby Ho atom. By fitting the known $1/r³$ law of the
dipolar interaction at different Fe-Ho distance, the authors
are able to measure the intrinsic magnetic moment of a single
Ho atom. This experiment shows that using quantum effects, energy scale is
in the sub $\mu$-eV range~\cite{Taeyoung_2017}.}
From measuring minute magnetic fields to having
access to the very-low energy scale of superconducting gaps, quantum
metrology can have strong impact in biosensors, industry and creating new
standards of measurements, as is already the case with the redefinition
of the SI unit system in 2018.

In summary, spin chains on solid surfaces have become accessible and
are very interesting objects of research. This new research field is rich and
lively due to the extraordinary prospects of the scanning tunnelling microscope. 
New developments
are further advancing the field. These developments include the
possibility to measure forces and currents concurrently, permitting the
analysis of magnetic structures with unprecedented accuracy. Furthermore,
the newly time-resolved technics are expanding \nico{our insight into the basic understanding and ability to manipulate spins on the atomic scale. 
These advances have enabled researchers to measure lifetimes
and coherence times of spins on surfaces with unprecedented accuracy. }We
are gathering new insight \nico{into} the dynamics of superposition states and
interactions at play.  We can now explore new phases of matter,
particularly the newly discovered topological phases. Spin chain research
will be a fundamental \nico{rich field for future exploration} in all of these topics.
